\documentclass[journal,a4paper]{IEEEtran}

\usepackage{graphicx}
\usepackage[T1]{fontenc}
\usepackage{enumerate}
\usepackage{amsmath,amssymb,latexsym,amsfonts}
\usepackage{cite}
\usepackage{color}

\usepackage{hyperref}
\hypersetup{
	colorlinks  = true, 
	urlcolor    = blue, 
	linkcolor   = blue, 
	citecolor   = blue 
}

\usepackage{gensymb}

\usepackage{subfigure}

\newcommand{\figref}[1]{\figurename~\ref{#1}}
\newcommand{\tableref}[1]{Table~\ref{#1}}
\newcommand{\secref}[1]{Section~\ref{#1}}

\newcommand{\Realset}{\mathbb{R}}

\newcommand{\sign}{\hspace{2pt}\mathrm{sign}}

\usepackage{ifpdf}
\ifpdf
\usepackage{epstopdf}
\DeclareGraphicsExtensions{.pdf,.eps,.png,.jpg,.mps,.bmp}
\else
\usepackage[dvips]{epsfig}
\fi

\begin{document}
	
	\title{\huge A Generalized Forced Oscillation Method for tuning Proportional Resonant Controllers}
	
\author{Charles~Lorenzini$^*$,~
	Luís~Fernando~Alves~Pereira,~
	and~Alexandre~Sanfelice~Bazanella,~\IEEEmembership{Senior~Member,~IEEE}
	\thanks{$^*$Corresponding author}
	\thanks{This work has been submitted to the IEEE for possible publication. Copyright may be transferred without notice, after which this version may no longer be accessible. This work was supported by the CAPES, Coordenação de Aperfeiçoamento de Pessoal de Nível Superior, and CNPq, Conselho Nacional de Desenvolvimento Científico e Tecnológico, Brazil.}
	\thanks{C. Lorenzini and L. F. A. Pereira are with the Graduate Program in Electrical Engineering (PPGEE),
	Federal University of Rio Grande do Sul, Porto Alegre 90035-190, Brazil (e-mail: charles.lorenzini@ufrgs.br; lfpereira@ufrgs.br)}
	\thanks{A. S. Bazanella is with the Department of Automation and Energy (DELAE),
	Federal University of Rio Grande do Sul, Porto Alegre 90035-190,
		Brazil (e-mail: bazanella@ufrgs.br)}}

	\maketitle
	
	\begin{abstract}
		This paper proposes a tuning methodology for proportional resonant (PR) controllers by using the design philosophy of the Ziegler-Nichols forced oscillation method. Unlike such related methods that are usual for PID design, and those that have been recently proposed for PR controllers, the method in this paper is not restricted to plants whose Nyquist diagram crosses the negative real axis. It involves a feedback experiment  with a relay of adjustable phase, which allows the identification of the most appropriate point of the frequency response for each class of plants. The validation of the proposed method, which includes the identification experiment and specific tuning formulas that provide appropriate stability margins for each class of plants, is performed by means of numerical examples in a large variety of plants, showing its wide applicability.
	\end{abstract}
	
	
	
	\section{Introduction}
	
	\IEEEPARstart{T}{he} internal model principle (IMP) \cite{art:contr:francis:1975} is a basic concept in control systems.
	It basically establishes that to asymptotically track a given reference, or to asymptotically reject a given disturbance, assuming a stable closed loop, the controller must have the nonvanishing modes of the reference and disturbance inputs. When these inputs are steps, the controller must have an integrator (a pole at the origin), like 
	the proportional-integral-differential (PID) controllers. 
	
	When these inputs are sinusoidal signals with frequency $\omega_r$, the internal mode is formed by a pair of poles at $\pm j\omega_r$. This characteristic leads to a controller with resonance frequency at $\omega_r$, hence the denomination of resonant controller.
	This kind of controller is widely applied in dc-ac inverters 
	\cite{art:res:per:2014:multiplos_ressonantes}\cite{art:res:Fukuda:2001:filtro_ativo}\cite{art:res:Gonzatti:2016:active_filter,art:res:Teodorescu:2006:ups}
	, high-precision positioning systems \cite{art:res:Habibullah:2017:vibracao}, vibration control in flexible structures \cite{art:res:Moheimani:2005:ress_struc}, and reduction of torque and flux ripples of permanent magnet synchronous machines \cite{art:res:Abosh:2017:torq}. In some applications, the so-called  high-Q resonant controllers are used, in which the controller poles are shifted to the left half of the complex plane, 
improving robustness at the cost of losing perfect tracking and rejection.

	In order to contribute to the applicability and enlarge the dissemination of resonant controllers, whose parameters are typically tuned considering the knowledge of the plant model, a tuning method of a resonant structure for plants that have an ultimate frequency (that is, whose Nyquist plot crosses the negative real
	axis) has been proposed in \cite{art:pereira:2015:PR-ZN}. This method can be implemented experimentally in a straightforward manner, through the same standard relay feedback experiment \cite{art:astrom:1984:rele} that is commonly applied to PID controllers. Then, the parameters of the resonant structure can be calculated from simple tuning formulas, similarly to the  Ziegler-Nichols (ZN)-like methods.
		
%
	
	On the other hand, a PID tuning method based on a modified relay feedback experiment has been proposed in \cite{art:bazanella:2017:PID-rele-foi} for  plants with relative degree larger than one. Since it can be applied
	to  plants that
	are not amenable to the application of the traditional ZN-like tuning methods -- plants that have
	neither an ultimate frequency nor an S-like reaction curve -- the class of plants for which ZN-like methods can be applied has been substantially enlarged.

	In this paper, a tuning method for proportional resonant  (PR) controllers is proposed that can be 
	applied to quite general linear time invariant plants, regardless of the existence of an ultimate frequency, of order
	or relative degree. This method is based on the modified relay feedback experiment presented in \cite{art:bazanella:2017:PID-rele-foi}, which enables the identification of a previously specified point of the plant's frequency response in a single experiment. Then, 
	similarly to \cite{art:pereira:2015:PR-ZN},  tuning formulas for the PR controller are developed
	to place this point of the loop frequency response at a specified location in the complex plane,
	chosen to provide appropriate stability margins. A detailed analysis and validation of the proposed method are performed in a wide variety of plants, which demonstrate its applicability to large amount of plants with different characteristics. Thus, this method provides an alternative to tune PR controllers without need of the plant model and with very little design effort.
	
	This paper is organized as follows. In \secref{sec:prel}, preliminary concepts are presented. The tuning method of the PR controller is proposed in \secref{sec:gfo}, where initially the control design philosophy is introduced, then the PR tuning formulas are developed based on the knowledge of the controller's resonance frequency and a particular point of the plant's frequency response. In \secref{sec:rap}, it is presented the modified relay feedback experiment that allows obtaining this information in a single experiment. The applicability of the proposed tuning methodology is verified in a wide variety of plants, a detailed analysis with three different plants is performed in \secref{sec:bench_plants}, and results obtained in a large range of plants are summarized in \secref{sec:test_batch}. Some concluding remarks are drawn in \secref{sec:conclusion}.


\section{Preliminaries}\label{sec:prel}


\subsection{Plants}

In this paper, linear time invariant causal (LTIC) plants are considered, which are represented by
\begin{equation}
\label{eq:TF_Ys}
Y(s) {} = {} G(s)U(s),
\end{equation}
where $U(s)$ and $Y(s)$ are the Laplace transforms of the input and the plant's output (the controlled variable), respectively, 
and $G(s)$ is the plant's transfer function, which is assumed to be strictly proper and BIBO-stable.
The plant is controlled with unitary feedback by
a LTIC controller, that is
\begin{equation}
\label{eq:TF_Es}
E(s) {} = {} R(s) - Y(s),\;\;\;U(s) {} = {} C(s)E(s)
\end{equation}
where $R(s)$ is the reference, $E(s)$ is the tracking error, and $C(s)$ is the controller's transfer function.



\subsection{PR Controller}

The controllers considered in this paper are Proportional-Resonant (PR) controllers in the form
\begin{equation}
\label{eq:C_pr_xi}
C_{pr}(s) {} = {} K_p + \frac{K_{r_1} s + K_{r_2}}{s^2+2\xi\omega_r s+\omega_r^2}
\end{equation}
where $\omega_r$ is the frequency that must be tracked and/or rejected, $\xi$ is the damping coefficient of the poles, and  $K_p$, $K_{r_1}$, $K_{r_2}\in \Realset$ are the gains to be tuned. 

A stable closed-loop system asymptotically tracks/rejects a given sinusoidal signal with frequency $\omega_r$ if the internal
mode formed by a pair of poles at $\pm j\omega_r$ is present in the loop, providing infinite gain at this frequency;
this is achieved by the PR controller \eqref{eq:C_pr_xi} with damping coefficient $\xi = 0$.
In many situations positive values of the damping coefficient are preferred \cite{art:res:Teodorescu:2006:ups}\cite{art:res:Habibullah:2017:vibracao}\cite{art:res:Moheimani:2005:ress_struc}\cite{art:res:Abosh:2017:torq},
which improves robustness and makes the tuning task easier at the cost of losing perfect tracking.
In this paper we develop a tuning method for the general case - i.e. arbitrary $\xi\geq0$. 


\subsection{Time-Domain Performance Assessment}

To assess the control systems in a systematic and standard-compliant way, performance criteria are 
evaluated in terms of the system's response to a sinusoidal reference signal: the settling time $t_s$ and the maximum overshoot $M_o$, which are defined as follows \cite{art:pereira:2015:PR-ZN}.

Let the reference be a sinusoidal signal, that is
\begin{equation}
\label{eq:sen_r}
r(t) = 0 \;\;\; \forall\, t < 0,\;\;\; r(t) = a_r \sin\left(\omega_r t\right) \;\;\;\forall \,t> 0 
\end{equation}
for some given amplitude $a_r>0$ and frequency $\omega_r \in \Realset$. The settling time is defined as the smallest time for the normalized tracking error reach and stay within a user-defined tolerance, that is
\begin{equation}
\label{eq:ts}
t_s = \min_{t_1} \;: \; \left|\frac{e(t)}{a_r}\right| < \epsilon \;\;\;\forall \, t>t_1, 
\end{equation}
where $\epsilon$ is the user-defined tolerance - usually $\epsilon\in [0.02;\,0,05]$ and in this paper is defined as $\epsilon = 0.02$. Besides that, it may be appropriate to represent the settling time in number of periods of the reference instead of (or in addition to) time units, i.e., to use the following measure:
\begin{equation}
\label{eq:ns}
n_s = \frac{\omega_r t_s}{2 \pi}.
\end{equation}
The maximum overshoot is the amount the plant's output exceeds its steady-state value and it is defined as
\begin{equation}
\label{eq:Mo_s}
M_o = \max \left\{\frac{y_{max} - y_r}{y_r}, 0 \right\}\times 100,
\end{equation}
where $y_{max} = \max_{t<t_s} \left|y\right(t\left)\right|$ is the maximum absolute value during transient response and $y_{r} = \max_{t>t_s} \left|y\right(t\left)\right|$ is the maximum absolute value of $y(t)$ in steady-state.



\subsection{Tuning methods based on forced oscillation} \label{sec:CFO}

There is a large amount of literature dedicated to PID tuning rules, and several methods have been proposed and successfully applied
since the seminal work \cite{art:pid:ZN:1942} -- an overview is given in \cite{book:pid:astrom:1995:pid}.
Many of these methods constitute variations to the closed-loop method originally presented in \cite{art:pid:ZN:1942}, which 
in this paper is referred to as the \textit{classical forced oscillation (CFO)} method.  

The CFO method is based on the knowledge of the \textit{ultimate point} of the plant's frequency response. The ultimate point of a given transfer function is the point at which its Nyquist plot crosses the negative real axis -- the point corresponding to the lowest frequency where its phase is $-\pi$. This point is characterized by the ultimate frequency $\omega_u$ and the ultimate gain $K_u$, which are defined as
\begin{equation}
\label{eq:crit_point}
\begin{aligned}
\omega_u {} = {} \underset{\omega \geq 0}{\textnormal{min}}\; \omega : \angle G(j\omega) = -\pi,\;\;
K_u   {} = {}\frac{1}{|G(j\omega_u)|}.
\end{aligned}
\end{equation}
These definitions allow to summarize the CFO method as follows.
\begin{enumerate}
	\item Identify the ultimate point of the plant's frequency response, that is, determine $\omega_u$ and $K_u$.
	\item Design the parameters of the controller such that $C(j\omega_u)G(j\omega_u) = p$, or equivalently
	\begin{equation}
	\label{eq:cont_zn}
	C(j\omega_u) = - K_u p 
	\end{equation}
	where $p$ is a prespecified location in the complex plane.
\end{enumerate}

The first step of the method is usually performed by means of the relay feedback experiment, which consists in a closed-loop experiment with the following nonlinear control action:
\begin{equation}
\label{eq:u_relay}
u(t) = d \sign(e(t)) + b,
\end{equation}
where $\sign(\cdot)$ is the sign function [$\!\sign(x)=1$ for positive $x$ and $\sign(x)=-1$ for negative $x$], $d\in\Realset^+$ is a parameter to be chosen, and $b\in\Realset$ is the bias. The parameter $d$ regulates the oscillation amplitude at the plant output and $b$ must be adjusted to obtain a symmetrical oscillation. Once a symmetric oscillation is obtained, its amplitude $A_u$ and period $T_u$ are measured and the ultimate quantities are calculated from \cite{art:astrom:1984:rele}
\begin{equation}
\label{eq:K_u-w_u-realy}
K_u = \frac{4d}{\pi A_u}\;\;\;\textnormal{and}\;\;\; \omega_u = \frac{2\pi}{T_u}.
\end{equation}

The second step of the method is fulfilled by solving \eqref{eq:cont_zn} for the controller's gains $K_p$, $T_i$ and $T_d$ with a chosen location $p$. Under the reasonable assumption that the frequency response of the plant is sufficiently smooth, shifting the ultimate point away from $-1$ in the complex plane implies that the whole open-loop frequency response is shifted away from it, thus leading to good stability margins. Over the years, different locations $p$ have been proposed, each one resulting in different transient performance and stability margins. The tuning formulas presented in \cite{art:pid:ZN:1942} correspond to $p = -0.4+j0.08$ for PI controllers and $p = -0.6 -j0.28$ for PID controllers.

Plants that have no ultimate point are not amenable to the application of the CFO method. This is the case of all the minimum-phase stable second-order plants and most plants with relative degree smaller than three, for instance. In order to overcome this limitation,
a PID tuning method based on a modified relay feedback experiment was presented in \cite{art:bazanella:2017:PID-rele-foi} that
is applicable to a large class of plants that have no ultimate point - the Extended Forced Oscillation (EFO) method.
The EFO can be applied to plants with relative degree larger than one and requires only one experiment without designer
intervention, keeping the same simplicity of the CFO method.

On the other hand, these methods were classically limited to the tuning of PID controllers.
Regarding the tuning of PR controllers, a method based on the CFO method has been proposed in \cite{art:pereira:2015:PR-ZN}, 
where tuning formulas were developed by solving \eqref{eq:cont_zn} for a given resonant structure. Thus the controller parameters can be obtained experimentally in a straightforward manner through the relay feedback experiment and simple tuning formulas, which depend on the plant's ultimate point and the frequency that must be followed and/or rejected.  Still, the tuning method and
resulting tuning formulas were restricted to plants whose frequency response possesses an ultimate point.

In the following Section we present a tuning method for PR controllers which is developed in the same spirit of the methods 
mentioned previously and that can be applied to quite generic plants, whether or not with an ultimate point, even of first order.



\section{Generalized Forced Oscillation Method}\label{sec:gfo}


In the same spirit of the CFO and related methods, the first step of the 
Generalized Forced Oscillation (GFO) method consists in identifying the point of the plant's frequency response at which 
the phase reaches a previously specified value $\nu$. That is, determine the quantities $\omega_\nu$ and 
$M_\nu$ defined as
	\begin{equation}
	\label{eq:Gwv}
	\omega_\nu = \underset{\omega \geq 0}{\textnormal{min}}\; \omega : \angle G(j\omega) = \nu \;\;\; \textnormal{and}\;\;\;M_\nu = |G(j\omega_\nu)|,
	\end{equation}
	which can also be written as
	\begin{equation}
	\label{eq:Gwv_res}
	G(j\omega_{\nu}) = M_\nu \angle \nu = M_\nu \left(\cos\left(\nu\right) + j\sin\left(\nu\right)\right).
	\end{equation}
In \secref{sec:rap} the modified relay feedback experiment that yields $\omega_\nu$ and $M_\nu$ will be presented. For the moment assume that these quantities have somehow become available. 
Then, design the controller parameters such that
\begin{equation*}
\label{eq:CGwv_p_gene}
C(j\omega_{\nu})G(j\omega_{\nu}) = p = M_\rho \left(\cos\left(\rho\right) + j\sin\left(\rho\right)\right),
\end{equation*}	
or equivalently,
\begin{equation}
\label{eq:Cwv_p_gene}
C(j\omega_{\nu}) = \frac{M_\rho}{M_\nu} \left(\cos(\rho-\nu) + j\sin(\rho-\nu)\right),
\end{equation}	
where $p$ is a previously specified location in the complex plane.

In what follows, PR tuning formulas will be proposed that correspond to  the solution of \eqref{eq:Cwv_p_gene} for the controller transfer function \eqref{eq:C_pr_xi} with $s = j\omega_{\nu}$. We restrict our presentation to the case where $\omega_r<\omega_\nu$.

\subsection{Defining $\nu$ and $p$}

To proceed with the design, it is necessary to define what point of the plant's frequency response will be identified, which 
corresponds to defining which value of $\nu$ to use. This must be a value such that the identified point is relevant and representative
of the stability margins that will be obtained. Then, it remains to specify what would be a reasonable phase margin to achieve,
which corresponds to defining the value of $p$. Control textbooks suggest values of phase margin of at least  $45\degree$ to
provide appropriate robustness and dynamic performance for typical practical situations \cite{book:wolovich:1993:automatic}.

To make these definitions, PR controllers have been designed for a wide array of plants, considering different identified points of the
plant's frequency response, and also with different specifications of phase margin. Several values of $\nu$ and $p$, which result in different close-loop performance and stability margins, have been tested. Then, the resulting closed-loop performance from the response to a sinusoidal reference signal has been evaluated. The best values of $\nu$ and $p$ for each class of plants are presented next.

For plants with an ultimate point, this is clearly  the point that must be used, so the choice $\nu=-180^o$ is self-evident
for this class of plants - which for future convenience we will call Class A.
In this particular case, \eqref{eq:Cwv_p_gene} can be rewritten as \eqref{eq:cont_zn}. 
Thus, for the plants in Class A, the ultimate point of the plant's frequency response is identified, that is, 
\begin{equation}
\label{eq:Gw_180}
\begin{aligned}
&\nu = -180\degree,\; \omega_\nu = \omega_u = \underset{\omega \geq 0}{\textnormal{min}}\,\omega : \angle G(j\omega) = -180\degree,\\
&M_\nu = M_u  = |G(j\omega_u)| = 1/K_u. 
\end{aligned}
\end{equation}
As for the choice of the location to which the ultimate point should be shifted, we took  
the classical Ziegler-Nichols point for PI tuning ($p=0.4+\jmath 0.08$) as a first approximation and,
after numerous tests in numerous plants, we have found that the following choice of $p$ provides the
best results:
\begin{equation}
\label{eq:p_a_m3}
p = 0.4\left(\cos(-183\degree)+j \sin(-183\degree)\right), 
\end{equation}
for $0<\omega_r/\omega_u<0.5$, and
\begin{equation}
\label{eq:p_a_m1}
p = 0.4\left(\cos(-181\degree)+j \sin(-181\degree)\right) 
\end{equation}
for $0.5 \leq \omega_r/\omega_u < 1$.


Consider now the plants that do not possess an ultimate point. 
In previous work \cite{art:bazanella:2017:PID-rele-foi} it was found that $\nu=-120^o$
was the best choice for the tuning of PID controllers
for plants without an ultimate point. As for $p$, it has been proposed in  \cite{art:bazanella:2017:PID-rele-foi}  to pick it such that a phase margin
of $50^o$ was achieved, which corresponds to $p = 1 \angle -130^o$. We have successfully tested these same
choices here for the tuning of PR controllers, so this is what  we propose, provided that the plant's 
frequency response achieves this phase for some frequency. This set of plants - that is, those that do not possess
an ultimate point but whose frequency response reaches $-120^o$ for some frequency - will be called class B in this paper.
Thus, for plants in Class B we propose to use 
\begin{equation}
\label{eq:Gw_120}
\begin{aligned}
&\nu = -120\degree,\,
\omega_\nu = \omega_{120} = \underset{\omega \geq 0}{\textnormal{min}}\, \omega \!:\! \angle G(j\omega) = -120\degree,\\
&M_\nu = M_{120} = |G(j\omega_{120})|
\end{aligned}
\end{equation}
and \begin{equation}
\label{eq:p_pr_b}
p = 1 \angle -130^o = \cos(-130\degree) + j\sin(-130\degree).
\end{equation}
With these definitions the phase margin will be $50\degree\!$, provided that the magnitude of the loop transfer function monotonically decreases for frequencies higher than $\omega_{120}$.

Finally, consider those plants whose frequency response never reaches $-120^o$, which will be
called Class C in this paper. For them a different value of $\nu$, and thus also of $p$, must be used.
This class of plants are the least problematic regarding stability margins, since the loop frequency
response of such a plant with a PR controller  is very unlikely to ever cross the negative real axis.
So, the choice of the design parameters $\nu$ and $p$ is not expected to be critical in this case;
still, the best choices must be made. After frequency response analysis of the problem
and a large number of tests, the best results were obtained with the following values,
which are thus the ones recommended for Class C: 
\begin{equation}
\label{eq:Gw_60}
\begin{aligned}
&\nu = -60\degree,\,
\omega_\nu = \omega_{60} = \underset{\omega \geq 0}{\textnormal{min}}\, \omega \!:\! \angle G(j\omega) = -60\degree,\\
&M_\nu = M_{60} = |G(j\omega_{60})|
\end{aligned}
\end{equation}
and \begin{equation}
\label{eq:p_pr_c}
p = 1 \angle -90\degree = \cos(-90\degree) + j\sin(-90\degree),
\end{equation}
which provide a phase margin of $90\degree$, assuming the monotonically decrease
of the loop transfer function for frequencies higher than $\omega_{60}$.

In the following, a set of generic tuning formulas is developed from the solution of \eqref{eq:Cwv_p_gene} with the proposed PR structure. Then,  particular sets of PR tuning formulas are proposed for each of the classes of plants previously defined.

\subsection{PR Tuning}

The PR tuning formulas are obtained by substituting \eqref{eq:C_pr_xi} with $s = j \omega_\nu$ into the tuning equation \eqref{eq:Cwv_p_gene}, which results in 
\begin{equation*}
\label{eq:Cpr1_wu_Kup_xi}
K_p \!+\! \frac{j \omega_{\nu} K_{r_1} + K_{r_2}}{\omega_r^2\!-\!\omega_{\nu}^2\!+\!j 2\xi\omega_r \omega_{\nu}} = {}
\frac{M_\rho}{M_\nu} (\cos(\rho-\nu)+ j \sin(\rho-\nu)).
\end{equation*}
After simplifications, the last equation can be expressed as
\begin{equation}
\label{eq:Cpr1_wu_Kup_3}
\begin{aligned}
& K_{p}\left(\omega_{r}^2 -\omega_\nu^2\right) + K_{r_2} + j \omega_\nu\left( K_{r_1} + 2K_p\xi\omega_r\right) = \\
& \frac{M_\rho}{M_\nu} \left( \left(\omega_{r}^2 -\omega_\nu^2\right) \cos(\rho-\nu) - 2\omega_\nu \xi\omega_r \sin(\rho-\nu)\right) +\\ 
& j \frac{M_\rho}{M_\nu} \left(\left(\omega_{r}^2 -\omega_\nu^2\right) \sin(\rho-\nu) + 2 \omega_\nu \xi\omega_r \cos(\rho-\nu)\right).
\end{aligned}
\end{equation}

It follows from real and imaginary parts of \eqref{eq:Cpr1_wu_Kup_3} that 
\begin{equation}
\label{eq:Cpr1_Re_xi}
\begin{aligned}
& K_{p}\left(\omega_{r}^2 -\omega_\nu^2\right) + K_{r_2} = \\ 
&\frac{M_\rho}{M_\nu} \left( \left(\omega_{r}^2 -\omega_\nu^2\right) \cos(\rho-\nu) - 2\omega_\nu \xi\omega_r \sin(\rho-\nu)\right),
\end{aligned}
\end{equation}
\begin{equation}
\label{eq:Cpr1_Imag_xi}
\begin{aligned}
& \omega_\nu\left( K_{r_1} + 2K_p\xi\omega_r \right) = \\
&\frac{M_\rho}{M_\nu} \left(\left(\omega_{r}^2 -\omega_\nu^2\right)\sin(\rho-\nu) +  2 \omega_\nu \xi\omega_r \cos(\rho-\nu)\right).
\end{aligned} 
\end{equation}

The expression in \eqref{eq:Cpr1_Re_xi} can be rewritten as
\begin{equation}
\label{eq:Cpr1_Kp_K2_xi}
\begin{aligned}
K_{p} {} = {} & \frac{K_{r_2}}{\omega_\nu^2 - \omega_{r}^2} + \frac{M_\rho \left(\omega_{r}^2 -\omega_\nu^2\right) \cos(\rho-\nu)}{M_\nu \left(\omega_{r}^2 -\omega_\nu^2\right)}-\\
&{} - {} \frac{2 M_\rho \omega_\nu \xi\omega_r \sin(\rho-\nu)}{M_\nu \left(\omega_{r}^2 -\omega_\nu^2\right)},
\end{aligned}
\end{equation}
where a degree of freedom in the tuning of the controller's parameter is verified, since this is a single equation with two unknowns: $K_p$ and $K_{r_2}$. A second equation involving these unknowns is obtained by imposing the product of the controller zeros to be equal to $\eta^2\omega_{r}^2$, where $\eta\!>\!0$ is a parameter to be determined next. This additional constraint allows to achieve controller zeros with absolute value $\eta\omega_r$ when they are complex conjugate, as detailed in the following.

The transfer function of the PR controller \eqref{eq:C_pr_xi} can be represented as
\begin{equation}
\label{eq:Cpr1_b_xi}
C_{pr}(s){} = {} K_{p} \dfrac{s^2 + \left(2\xi\omega_r + \frac{K_{r_1}}{K_{p}} \right)\!s + \left(\frac{K_{r_2}}{\omega_{r}^2K_{p}}+1\right)\!\omega_{r}^2}
{s^2+2\xi\omega_r s + \omega_r^2}.
\end{equation}

The numerator of $C_{pr}(s)$ expressed as a function of $K_p$ and its roots, say $z_1$ and $z_2$, is given by 
\begin{equation*}
\label{eq:Cpr1_z1_xi}
Z(s) = K_p(s-z_1)(s-z_2) = K_p\left( s^2 - (z_1 + z_2)s + z_1 z_2\right).
\end{equation*}
From the additional constraint $z_1 z_2 = \eta^2 \omega_r$, if the controller has a pair of complex zeros, i.e. $\!z_{1,2} = -a \pm jb$, where $a,b\in\Realset$, then $z_1 z_2 = a^2 + b^2 = \eta^2 \omega_r^2$ and $|z_{1,2}| = \eta \omega_r$. Thus, the numerator of $C_{pr}(s)$ can be rewritten as
\begin{equation}
\label{eq:Cpr1_z2_xi}
Z(s) = K_p\left(s^2 + 2as + \eta^2 \omega_r^2\right).
\end{equation}

By comparing the expressions in \eqref{eq:Cpr1_b_xi} and \eqref{eq:Cpr1_z2_xi}, a second equation involving the parameters $K_p$ e $K_{r_2}$ is achieved
\begin{equation}
\label{eq:Cpr1_raio_zeros2_xi}
K_{p} = \frac{K_{r_2}}{\left(\eta^2-1\right) \omega_{r}^2}.
\end{equation}

From \eqref{eq:Cpr1_Kp_K2_xi} and \eqref{eq:Cpr1_raio_zeros2_xi}, the generic tuning formula for $K_2$ is obtained
\begin{equation}
\label{eq:Cpr1_K2_xi}
\begin{aligned}
K_{r_2} {} = {} & \frac{M_\rho \left(\omega_{r}^2 -\omega_\nu^2\right) \cos(\rho-\nu)\left(\eta^2-1\right) \omega_{r}^2}{M_\nu \left(\eta^2\omega_{r}^2  -\omega_\nu^2 \right)} -\\ &-\frac{2 M_\rho \omega_\nu \xi\omega_r^3 \sin(\rho-\nu)\left(\eta^2-1\right)}{M_\nu \left(\eta^2\omega_{r}^2  -\omega_\nu^2 \right)}.
\end{aligned}
\end{equation}
Substitution of the previous equation into \eqref{eq:Cpr1_raio_zeros2_xi} gives the generic tuning formula for $K_p$
\begin{equation}
\label{eq:Cpr1_Kp_xi}
K_{p} {}= {}\frac{M_\rho\! \left(\left(\omega_{r}^2 -\omega_\nu^2\right)\! \cos(\rho-\nu) \!-\! 2\omega_\nu \xi\omega_r \sin(\rho-\nu)\right)\!}{M_\nu \left(\eta^2\omega_{r}^2 -\omega_\nu^2 \right)}\!.
\end{equation}
The generic tuning formula for $K_{r_1}$ is obtained by substituting the last expression into \eqref{eq:Cpr1_Imag_xi}. Then, after simplifications, it becomes  the following expression.
\begin{equation}
\label{eq:Cpr1_K1_xi}
\begin{aligned}
K_{r_1} {} = {}  &\frac{M_\rho 2 \omega_\nu \xi \omega_r^3 \left(\eta^2 -1\right)\cos(\rho-\nu)}{{M_\nu\omega_\nu} \left(\eta^2\omega_{r}^2  -\omega_\nu^2 \right)} +\\
+ &  \frac{M_\rho \!\left(\left(\omega_{r}^2 -\omega_\nu^2\right)\!\left(\eta^2\omega_{r}^2  -\omega_\nu^2 \right) \!+\! 4\xi^2\omega_u^2\omega_{r}^2\right)\sin(\rho-\nu) }{{M_\nu\omega_\nu} \left(\eta^2\omega_{r}^2  -\omega_\nu^2 \right)}\!        
\end{aligned}
\end{equation}

After developing the set of generic tuning formulas presented in the equations \eqref{eq:Cpr1_K2_xi} to \eqref{eq:Cpr1_K1_xi}, sets of particular tuning formulas are obtained for each of the classes of plants previously defined. As mentioned before, this task was performed by considering a wide array of plants and different identified point of the plant's frequency response (for those without ultimate point), for several choices of locations $p$, damping coefficient $\xi$, and parameter $\eta$. 

The parameter $\eta$ was set to $0.1$ in order to achieve controller zeros (when they are complex) a decade below than the controller poles. The damping coefficient $\xi $ is a parameter to be chosen by the controller designer, considering that from the IMP a stable closed-loop system asymptotically tracks a given sinusoidal reference, or asymptotically rejects a given sinusoidal disturbance, with frequency $\omega_r$ if the PR controller \eqref{eq:C_pr_xi} with $\xi = 0$ is inserted in the loop.

For the Class A, the identified point of the plant's frequency response is given in \eqref{eq:Gw_180} and there are two locations $p$ depending on the relationship between $\omega_r$ and $\omega_u$. Consequently, two sets of tuning formulas of the PR controller \eqref{eq:C_pr_xi} are proposed considering \eqref{eq:p_a_m3} for $0<\omega_r/\omega_u<0.5$ and \eqref{eq:p_a_m1} for $0.5 \leq \omega_r/\omega_u < 1$. Thus, the set of tuning formulas of the PR controller for $0<\omega_r/\omega_u<0.5$ is:
\begin{equation}
\label{eq:Cpr_class_a_xi_m1}
\begin{aligned}
K_{p} {} = &{} \frac{0.4 \left(\omega_u^2 - \omega_{r}^2\right) - 0.0140 \omega_u \xi\omega_r}{M_u \left(\omega_u^2 - 0.01\omega_{r}^2\right)}\\
K_{r_1} {} = &{}  \frac{0.00698\left(\omega_u^2 - \omega_{r}^2\right)}{{M_u\omega_u} }+ \frac{0.0279\xi^2\omega_u^2\omega_r^2  + 0.792 \omega_u \xi \omega_r^3}{{M_u\omega_u} \left(\omega_u^2 - 0.01\omega_{r}^2\right)}\\
K_{r_2} {} = & {}  \frac{0.396\left(\omega_{r}^2 -\omega_u^2\right) \omega_{r}^2 + 0.0138  \omega_u \xi\omega_r^3}{M_u \left(\omega_u^2 - 0.01\omega_{r}^2\right)} \\ 
\end{aligned}, 
\end{equation}
whereas 
\begin{equation}
\label{eq:Cpr_class_a_xi_m3}
\begin{aligned}
K_{p} {} = &{} \frac{0.399 \left(\omega_u^2 - \omega_{r}^2\right) - 0.0419 \omega_u \xi\omega_r}{M_u \left(\omega_u^2 - 0.01\omega_{r}^2\right)}\\
K_{r_1} {} = &{}  \frac{0.0209\left(\omega_u^2 - \omega_{r}^2\right)}{{M_u\omega_u} }+ \frac{0.0837\xi^2\omega_u^2\omega_r^2  + 0.791 \omega_u \xi \omega_r^3}{{M_u\omega_u} \left(\omega_u^2 - 0.01\omega_{r}^2\right)}\\
K_{r_2} {} = & {}  \frac{0.395\left(\omega_{r}^2 -\omega_u^2\right) \omega_{r}^2 + 0.0414  \omega_u \xi\omega_r^3}{M_u \left(\omega_u^2 - 0.01\omega_{r}^2\right)} \\ 
\end{aligned}
\end{equation}
must be applied when $0.5 \leq \omega_r/\omega_u < 1$.

For the Class B, the identified point of the plant's frequency response is given in \eqref{eq:Gw_120} and the location $p$ is presented in \eqref{eq:p_pr_b}. Hence, the set of tuning formulas of the PR controller for a Class B plant is:
\begin{equation}
\label{eq:Cpr_class_b_xi}
\begin{aligned}
K_{p} {} = {}& \frac{0.985\left(\omega_{120}^2 - \omega_{r}^2\right) -0.347\omega_{120} \xi\omega_r}{M_{120} \left(\omega_{120}^2 - 0.01\omega_{r}^2\right)}\\
K_{r_1} {} = {}& \frac{ 0.174 \left(\omega_{120}^2 - \omega_{r}^2\right)}{M_{120}\omega_{120}} + \frac{0.695\xi^2\omega_{120}^2\omega_r^2 +1.95  \omega_{120} \xi \omega_r^3}{{M_{120}\omega_{120}} \left(\omega_{120}^2 -0.01\omega_{r}^2\right)}\\
K_{r_2} {} =  {}&  \frac{0.975 \left(\omega_{r}^2 -\omega_{120}^2\right)\omega_{r}^2 + 0.344 \omega_{120} \xi\omega_r^3}{M_{120} \left(\omega_{120}^2 - 0.01\omega_{r}^2\right)}
\end{aligned}
\end{equation}

For the Class C, the identified point of the plant's frequency response is given in \eqref{eq:Gw_60} and the location $p$ is defined in \eqref{eq:p_pr_c}. Therefore, the set of tuning formulas of the PR controller for a Class C plant is:

\begin{equation}
\label{eq:Cpr_class_c_xi_90}
\begin{aligned}
	K_{p} {} = {}& \frac{0.866 \left(\omega_{60}^2 - \omega_{r}^2\right) - \omega_{60} \xi\omega_r}{M_{60} \left(\omega_{60}^2 - 0.01\omega_{r}^2\right)},\\
	K_{r_1} {} = {}& \frac{0.5\left(\omega_{60}^2 - \omega_{r}^2\right)}{{M_{60}\omega_{60}}} + \frac{2 \xi^2\omega_{60}^2\omega_r^2 + 1.71 \omega_{60} \xi \omega_r^3}{{M_{60}\omega_{60}} \left(\omega_{60}^2 -0.01\omega_{r}^2\right)}\\
	K_{r_2} {} =  {}&  \frac{0.857 \left(\omega_{r}^2 -\omega_{60}^2\right) \omega_{r}^2 + 0.99  \omega_{60} \xi\omega_r^3}{M_{60} \left(\omega_{60}^2-0.01\omega_{r}^2 \right)}
\end{aligned}
\end{equation}

The PR controller applied to a Class A plant is tuned using two sets of tuning formulas because there is a compromise between the controller's contribution to the phase at the plant's ultimate frequency, stability margins, and closed-loop performance. Moreover, a small change in the controller's contribution to the phase at this specific frequency significantly changes the stability margins and the closed-loop performance. Unlike the Classes B and C for which the PR controller guarantees the defined phase margins of $50\degree$ and $90\degree$, respectively.

\begin{figure}[t!]
	\centering
	\includegraphics[width=0.90\columnwidth]{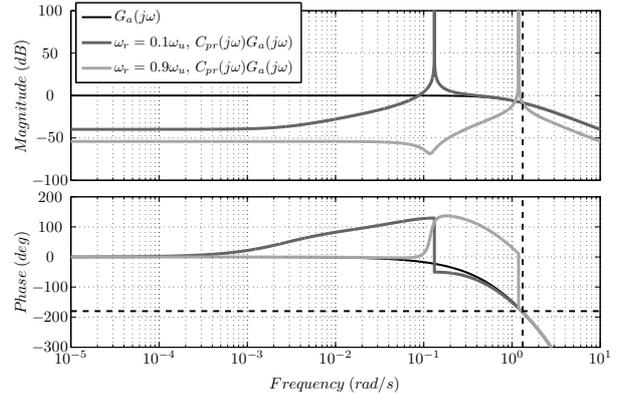}
	\caption{Frequency response of $G_a(j\omega)\! = \! e^{-s}/(s + 1)^2$ and $G_a(j\omega)C_{pr}(j\omega)$ with $\omega_r = 0.1\omega_u$ and $\omega_r = 0.9\omega_u$. Dashed lines are at $\omega_u$ and at $-180\degree$.}
	\label{fig:bode_G_14_C_wrs}
\end{figure}

A frequency response analysis helps to analyze the proposed PR tuning. \figref{fig:bode_G_14_C_wrs} presents the frequency response of the plant $G_a(j\omega)\! = \! e^{-s}/(s + 1)^2$ and of the loop $G_a(j\omega)C_{pr}(j\omega)$ with two reference frequencies: $\omega_r = 0.1\omega_u$ and $\omega_r = 0.9\omega_u$, for which the controllers are tuned through \eqref{eq:Cpr_class_a_xi_m1} and \eqref{eq:Cpr_class_a_xi_m3}, respectively. For $\omega_r = 0.1\omega_u$, the controller's resonance frequency is much lower than the plant's ultimate frequency, appropriate stability margins are achieved, and good closed-loop performance is expected. For $\omega_r = 0.9\omega_u$, the controller's resonance frequency is close to the plant's ultimate frequency, as consequence stability margins are much smaller than the previous case. Thus, poor performance and even closed-loop unstable systems are to be expected for $\omega_r \approx \omega_u$, since the controller structure presents very large gains in a range around the plant's ultimate frequency. 
It is important to notice that this is not a limitation of the tuning rules just proposed; it is rather a limitation of the controller structure. 

\begin{figure}[t!]
	\centering
	\includegraphics[width=0.90\columnwidth]{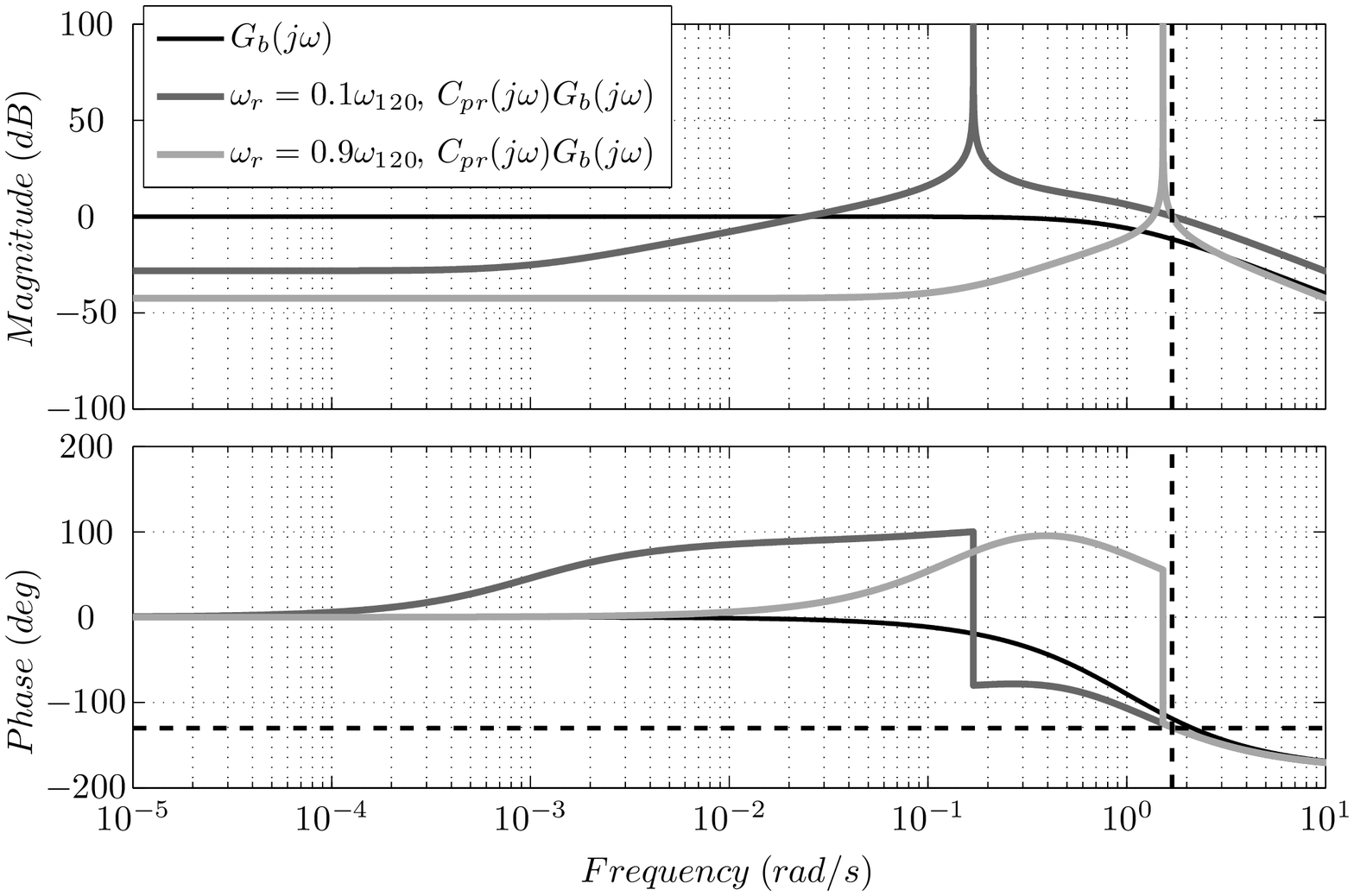}
	\caption{Frequency response of $G_b(j\omega) = 1/\left(s + 1\right)^2$ and $G_b(j\omega)C_{pr}(j\omega)$ with $\omega_r = 0.1\omega_{120}$ and $\omega_r = 0.9\omega_{120}$. Dashed lines are at $\omega_{120}$ and at $-130\degree$.}
	\label{fig:bode_G_2_C_wrs}
\end{figure}

\begin{figure}[t!]
	\centering
	\includegraphics[width=0.90\columnwidth]{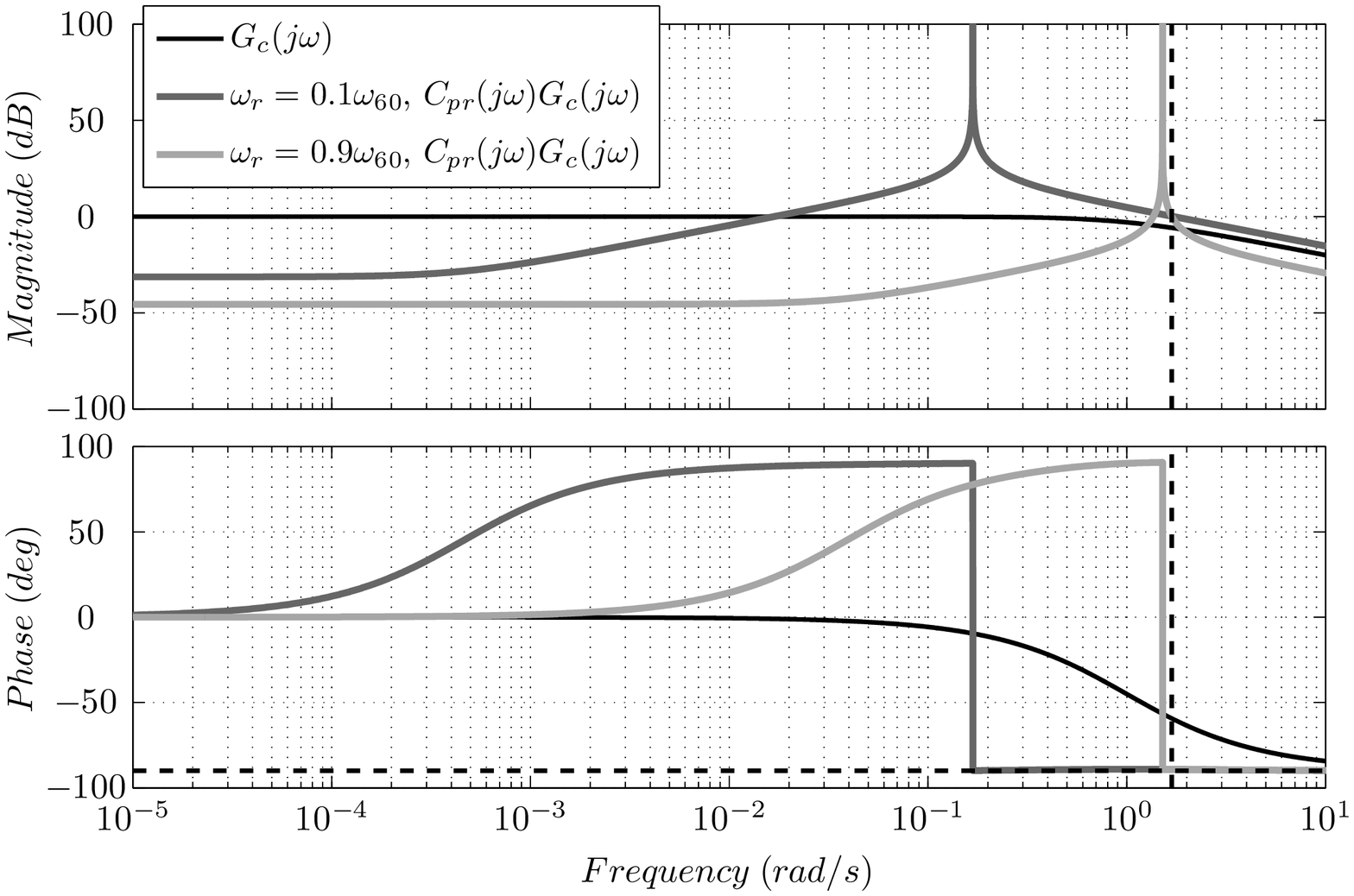}
	\caption{Frequency response of $G_c(j\omega) = 1/\left(s + 1\right)$ and $G_c(j\omega)C_{pr}(j\omega)$ with $\omega_r = 0.1\omega_{60}$ and $\omega_r = 0.9\omega_{60}$. Dashed lines are at $\omega_{60}$ and at $-90\degree$.}
	\label{fig:bode_G_1_C_wrs}
\end{figure}

	\figref{fig:bode_G_2_C_wrs} shows the frequency response of the Class B plant $G_b(j\omega) = 1/\left(s + 1\right)^2$ and of the loop $G_b(j\omega)C_{pr}(j\omega)$ with the PR controller tuned through \eqref{eq:Cpr_class_b_xi} having two reference frequencies: $\omega_r = 0.1\omega_{120}$ and $\omega_r = 0.9\omega_{120}$. In the same way, \figref{fig:bode_G_1_C_wrs} presents the frequency response of the Class C plant $G_c(j\omega) = 1/\left(s + 1\right)$ and of the loop $G_c(j\omega)C_{pr}(j\omega)$ with the PR controller tuned through \eqref{eq:Cpr_class_c_xi_90} having two reference frequencies: $\omega_r = 0.1\omega_{60}$ and $\omega_r = 0.9\omega_{60}$. For plants of the  Classes B and C, poor performance is also to be expected for $\omega_r \approx \omega_{\nu}$, but closed-loop unstable systems are not to be expected since the defined phase margin is guaranteed for $\omega_r < \omega_{\nu}$.




\section{Relay experiment with Adjustable Phase}\label{sec:rap}

The formulas for tuning of PR controllers presented in the previous Sections require knowledge of a previously specified
point of the plant's frequency response. In this Section a procedure to obtain this information  from a simple experiment is described.

\begin{figure}[t!]
	\centering
	\includegraphics[width=1\columnwidth]{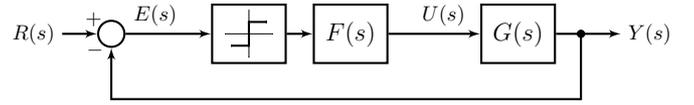}
	\caption{Relay experiment with Adjustable Phase (RAP experiment) for identification of the ultimate point of $F(s)G(s)$.}
	\label{fig:RAP}
\end{figure}

The relay feedback experiment described in \secref{sec:CFO} is a classical way to experimentally determine the ultimate point of a plant, but a slight change in this experiment allows to identify other points of the plant's frequency response \cite{book:pid:astrom:1995:pid}. When a known transfer function, say $F(s)$, is inserted in the loop in addition to the relay, as in \figref{fig:RAP}, if the self-oscillation behavior is obtained then it will have the ultimate frequency of the transfer function $F(s)G(s)$, that is, at $\omega_1 \!:\!\angle F(j\omega_1)G(j\omega_1) = -180\degree$.
Thus, the plant's magnitude and phase at this frequency can be calculated as \cite{art:bazanella:2017:PID-rele-foi}:
\begin{equation}
\label{eq:GFOI}
\left|G(j\omega_1)\right| {} = {} \frac{\pi A}{4 d \left|F(j\omega_1)\right|},\; \angle G(j\omega_1) {} = {} -180\degree - \angle F(j\omega_1),
\end{equation}
since $F\left(j\omega_1\right)$ is known.

In order to implement the proposed tuning method, it is necessary to identify the point of the plant's frequency response whose phase is $\nu$. If the identification is performed through the ultimate point of $F(s)G(s)$, a transfer function $F(s)$ whose phase is
\begin{equation}
\label{eq:gamma}
\gamma  \triangleq - 180\degree -\nu
\end{equation}
at the frequency $\omega_\nu$ must be chosen. But this frequency is not known in advance since it is one of the two quantities that the experiment aims at identifying.

To overcome this difficulty, the use of a transfer function $F(s)$ with (almost) constant phase - that is,
$\angle F(j\omega) = \gamma\;\forall \omega$ - has been proposed in \cite{art:bazanella:2017:PID-rele-foi}.
A system that has a transfer function with a flat phase frequency response that is not necessarily an entire multiple of
$-90\degree$ is an fractional order integrator (FOI) and it is represented by
\begin{equation}
\label{eq:F_s}
FOI(s) = \frac{1}{s^m},
\end{equation}
where it can be verified that
\begin{equation}
\label{eq:phase_F_s}
\angle FOI(j\omega) =  - \angle \left(\frac{j}{\omega}\right)^m  = - \angle \left(\frac{e^{\frac{j\pi}{2}}}{\omega}\right)^m= -\frac{\pi}{2}m.
\end{equation}
Defining $m = -\gamma/90\degree$ in \eqref{eq:phase_F_s} yields $\angle FOI(j\omega) = \gamma\; \forall \omega$ and, for example, choosing $\gamma = -30\degree$ results in $m = 1/3$ and for $\gamma = -60\degree$ is obtained $m = 2/3$.

Fractional-order systems are usually approximately implemented by integer order systems. To obtain transfer functions that approximate the magnitude and phase characteristics of a desired FOI, the MATLAB package FOMCON \cite{art:foi:tepljakov:2011:fomcon}, \cite{online:foi:Tepljakov:2016:FOI_site} was used. The transfer function presented in \eqref{eq:aFOI} with the two sets of coefficients shown in \tableref{tab:foi_30_60_tex} represents two FOIs approximations with magnitude characteristics of $-m\times20$ dB/decade and constant phase value of $-m\times90\degree$ for $m = 1/3$ and $2/3$, considering the range of frequencies from $10^{-3}$ to $10^3$ rad/s.
\begin{equation}
\label{eq:aFOI}
\hat F(s) = \frac{\sum_{k=0}^{11}b_k s^k}{\sum_{k=0}^{11}a_k s^k}
\end{equation}
\begin{table}[t!]
	\centering
	\renewcommand{\tabcolsep}{4pt}
	\caption{Coeficients of $\hat F(s)$}
	\begin{tabular}{c|c|c|c|c} 
		\hline
		& \multicolumn{2}{c|}{$m = 1/3$ for $\gamma = -30\degree$}&\multicolumn{2}{c}{$m = 2/3$ for $\gamma = -60\degree$}\\
		\hline
		$k$ & $a_k$ & $b_k$ & $a_k$ & $b_k$\\ 
		\hline
		0  & $ 0 $          & $ 0.3452 $      & $ 0 $          & $ 0.7152 $     \\ \hline 
		1  & $ 111.1 $      & $ 1309 $        & $ 11.11 $      & $ 1446 $       \\ \hline 
		2  & $ 8.49\times 10^4 $  & $ 5.4\times 10^5 $    & $ 1.097\times 10^4 $ & $ 4.387\times 10^5 $ \\ \hline 
		3  & $ 1.15\times 10^7 $  & $ 4.302\times 10^7 $  & $ 1.918\times 10^6 $ & $ 2.678\times 10^7 $ \\ \hline 
		4  & $ 3.232\times 10^8 $ & $ 7.22\times 10^8 $   & $ 6.963\times 10^7 $ & $ 3.473\times 10^8 $ \\ \hline 
		5  & $ 1.942\times 10^9 $ & $ 2.598\times 10^9 $  & $ 5.403\times 10^8 $ & $ 9.672\times 10^8 $ \\ \hline 
		6  & $ 2.509\times 10^9 $ & $ 2.013\times 10^9 $  & $ 9.016\times 10^8 $ & $ 5.799\times 10^8 $ \\ \hline 
		7  & $ 6.986\times 10^8 $ & $ 3.36\times 10^8 $   & $ 3.24\times 10^8 $  & $ 7.487\times 10^7 $ \\ \hline 
		8  & $ 4.195\times 10^7 $ & $ 1.211\times 10^7 $  & $ 2.506\times 10^7 $ & $ 2.08\times 10^6 $  \\ \hline 
		9  & $ 5.462\times 10^5 $ & $ 9.508\times 10^4 $  & $ 4.164\times 10^5 $ & $ 1.238\times 10^4 $ \\ \hline 
		10 & $ 1569 $       & $ 167.8 $       & $ 1466 $       & $ 15.45 $      \\ \hline 
		11 & $ 1 $          & $ 0.06905 $     & $ 1 $          & $ 0.003576 $   \\ \hline 
	\end{tabular}
	\label{tab:foi_30_60_tex}
\end{table}

\begin{figure}[t!]
	\centering
	\includegraphics[trim=0cm 0cm 0cm 0cm, clip=true, angle=0,width=0.95\columnwidth]{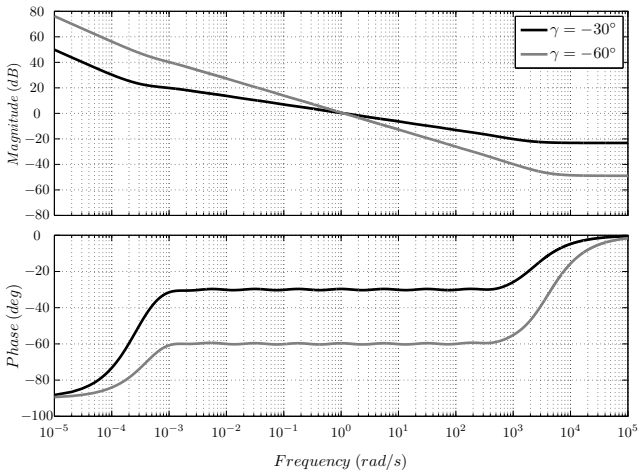}
	\caption{Frequency response of the FOI approximations.}
	\label{fig:plot_fois}
\end{figure}

\figref{fig:plot_fois} presents the frequency response of these two FOIs approximations with magnitude and phase curves having: $-6.66$ dB/decade and $-30 \degree$; $-13.33$ dB/decade and $-60 \degree$. On the other hand, a transfer function with constant phase
of $-120\degree$ can be obtained from the $-30\degree$ FOI
approximation by adding an integrator. These are the transfer functions we actually use in our design - that is, taking $F(s) = \hat F(s)$
with the parameters in the second column of Table \ref{tab:foi_30_60_tex}  and $F(s) = \hat F(s)/s$
with the parameters in the first column of Table \ref{tab:foi_30_60_tex}  . 

The desired point of the plant's frequency response is properly identified if $\hat F(s)$ has phase $\gamma$ in the self-oscillation frequency obtained in the relay feedback experiment with the approximated FOI. Thus, considering the frequency responses of $\hat F(j\omega)$ presented in \figref{fig:plot_fois}, the desired point of the plant's frequency response will be properly identified as long as 
$10^{-3}\leq\omega_{\nu}\leq10^{3}$ rad/s. Thus, care must be exercised in picking the range of frequencies for which
the integer dimension function $\hat F(s)$ properly approximates the FOI, which
is achieved by taking an approximation of sufficiently large order. 


To identify a particular point of the plant's frequency response, the proposed experiment employs an element that can be seen as a relay of adjustable constant phase for a defined range of frequencies. This experiment, whose block diagram is shown in \figref{fig:RAP}, was baptized \textit{Relay experiment with Adjustable Phase (RAP experiment)}, and it is described as follows.

The RAP experiment is started with a $0\degree$ phase relay, i.e., the traditional relay experiment is performed. If a self-oscillatory behavior is obtained and the plant's output is a well-defined signal, that is, its oscillation's amplitude and period are well defined, then, the plant will belong to the Class A, as stated  in \secref{sec:gfo}, and its ultimate point will be identified. If the self-oscillatory behavior is not obtained in the RAP experiment with $0\degree$, the plant has no ultimate point.

In this case, the relay's phase is decreased and the RAP experiment with $-60\degree$ is performed. If a self-oscillation condition is obtained, the point of the plant's frequency response whose phase is $-120\degree$ will be identified and the plant will belong to the Class B, as stated in \secref{sec:gfo}. If, once again, the self-oscillation condition is not achieved, the phase of the plant's frequency response does not reach either $-180\degree$ or $-120\degree$.

In this case, the relay's phase is decreased for the last time and the RAP experiment with $-120\degree$ is performed. If a self-oscillatory behavior is obtained, then the point of the plant's frequency response whose phase is $-60\degree$ will be identified, and the plant will belong to the Class C, as stated in \secref{sec:gfo}.

Thus, the RAP experiment enables an automatic procedure for identification of the relevant point of the plant's
frequency response -- whether having or not ultimate point -- in a single experiment without designer intervention. 
	

\section{Benchmark Class of plants}\label{sec:bench_plants}

In order to evaluate the proposed tuning methodology, three different plants will be considered. These plants represent cases of each of the three classes of plants considered in \secref{sec:gfo}. For each of them, a detailed analysis will be done and all steps of the controller design will be presented.


\subsection{Class A plant}\label{sec:plants_a}

The first plant considered is described by the following transfer function:
\begin{equation}
\label{eq:G_a}
G_{a}(s) {} = {} \frac{e^{-s}}{\left(s + 1\right)^2}.
\end{equation}

\begin{figure}[t!]
	\centering
	\includegraphics[width=0.9\columnwidth]{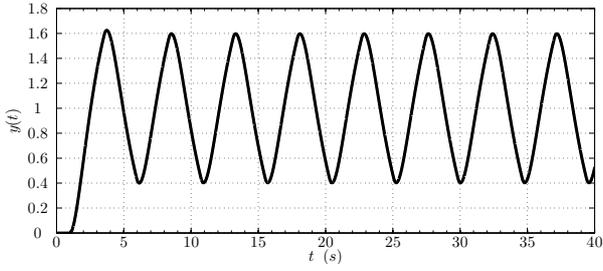}
	\caption{Closed loop response for the RAP experiment applied to the plant $G_a(s)$.}
	\label{fig:osc_G_14_tau_1}
\end{figure}

The starting point of tuning methodology is to identify a particular point of the plant's frequency response. Since the plant $G_{a}(s)$ has ultimate point it belongs to the Class A, then a self-oscillatory behavior is achieved in the RAP experiment with $0\degree$, and the plant's ultimate point is identified. The signal presented in \figref{fig:osc_G_14_tau_1} represents the plant's output in this experiment considering as reference input a step with amplitude one and the obtained parameters are summarized in \tableref{tab:resultados_G_14_FOI_0_Mu_Tu_tex}. 

\begin{table}[t!]
	\centering
	\renewcommand{\tabcolsep}{4pt}
	\caption{Parameters for the RAP experiment and $G_{a}(s)$}
	\label{tab:resultados_G_14_FOI_0_Mu_Tu_tex}
	\centering 
	\begin{tabular}{c|c|c|c|c|c|c|c} 
		\hline
		$d$   & $b$ & $A$     & $\gamma$   & $\nu$         & $|F(j\omega_{u})|$ & $M_{u}$ & $\omega_{u}\;(rad/s)$  \\ 
		\hline  
		$1.3$ & $1$ & $0.598$ & $0\degree$ & $-180\degree$ & $1$                & $0.391$ & $1.32$                 \\ \hline 
	\end{tabular} 
\end{table}

The second step is to design the PR controller \eqref{eq:C_pr_xi} for a given reference frequency $\omega_r$ using the identified point of the plant's frequency response. For a Class A plant, the controller gains are calculated depending on the relationship between $\omega_r$ and $\omega_u$: if $0<\omega_r/\omega_u<0.5$ the PR controllers must be tuned through \eqref{eq:Cpr_class_a_xi_m1}; whereas, if $0.5 \leq \omega_r/\omega_u < 1$ the PR controllers must be tuned through \eqref{eq:Cpr_class_a_xi_m3}.

\begin{figure}[t!]
	\centering
	\subfigure[$\omega_r = 0.1 \omega_u$]{\includegraphics[width=0.85\columnwidth]{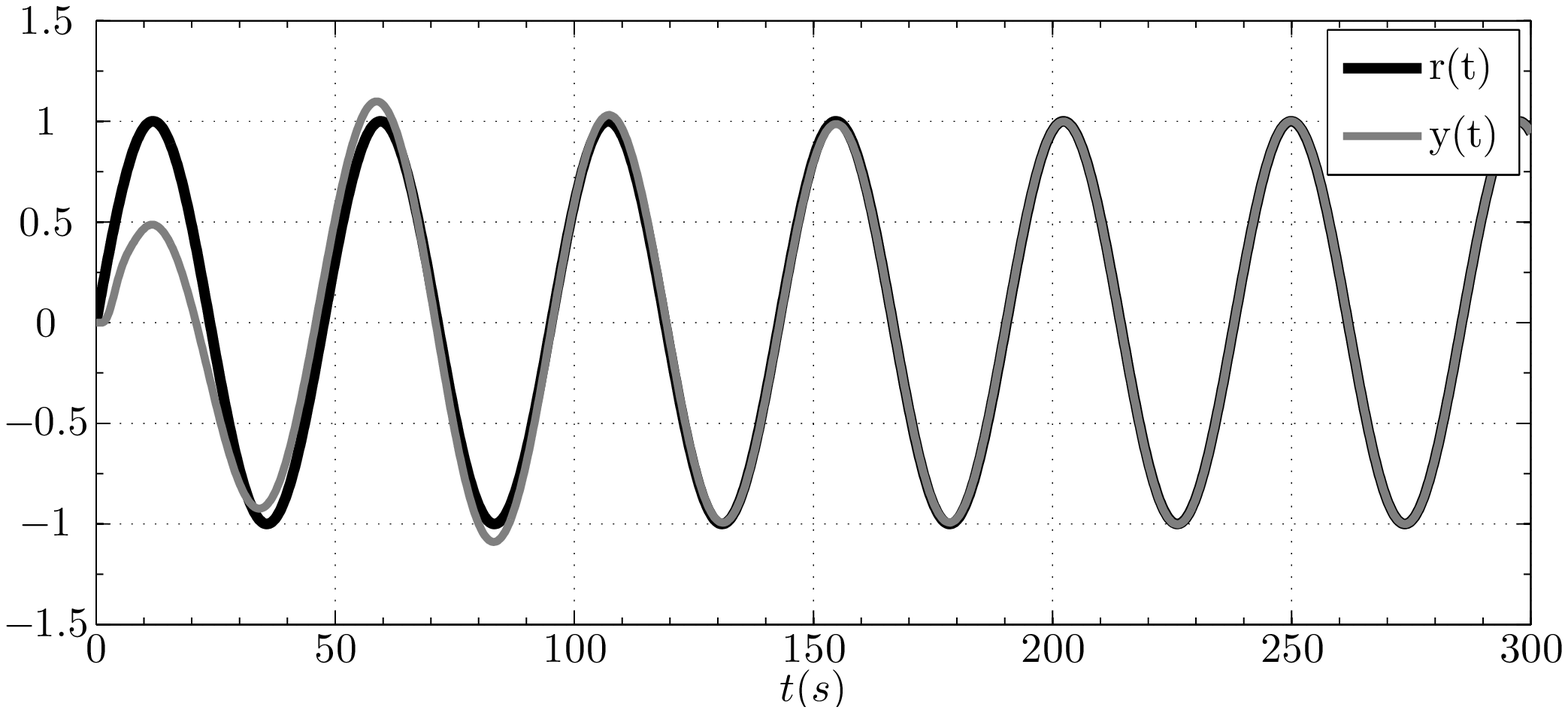}
		\label{fig:out_res_G14_w_0_1}}
	\vskip-0.20cm
	\subfigure[$\omega_r = 0.7 \omega_u$]{\includegraphics[width=0.85\columnwidth]{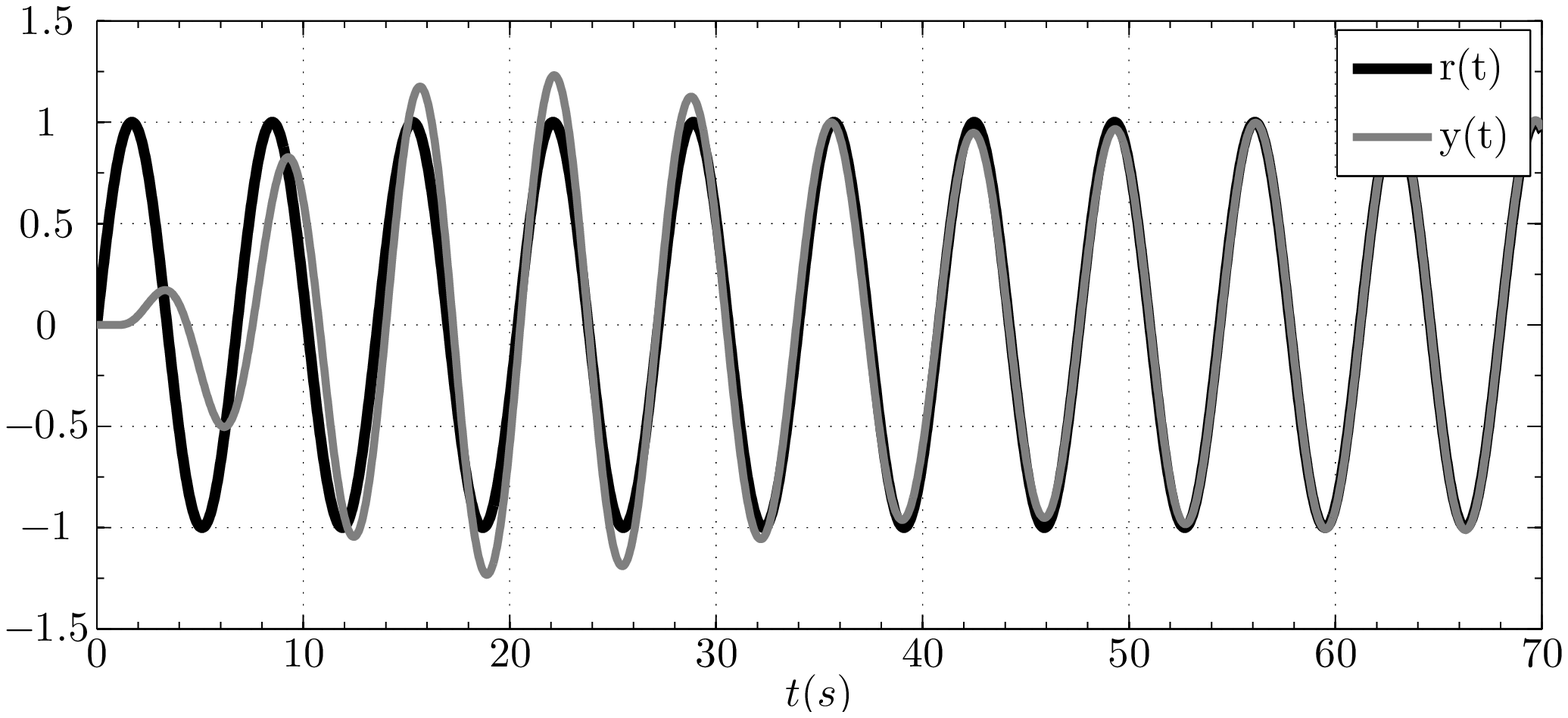}
		\label{fig:out_res_G14_w_0_7}}
	\caption{Closed-loop response of $G_a(s)$ with PR controller.}
	\label{fig:out_res_G14}
\end{figure}

In this numerical example, the reference is a sinusoidal signal with frequencies $\omega_r = 0.1\,\omega_{u}$ and $\omega_r = 0.7\,\omega_{u}$.  
The damping coefficient $\xi$ is chosen equal to zero in order to achieve asymptotically tracking of the reference signal. The set of controller's parameters and performance measures are summarized in \tableref{tab:resultados_G_14_FOI_0_tex}. The reference and the output signals for each set of controller's parameters are shown in \figref{fig:out_res_G14}.

\begin{table}[t!]
	\centering
	\renewcommand{\tabcolsep}{3pt}
	\caption{Tuning parameters and performance measures for $G_{a}(s)$}
	\label{tab:resultados_G_14_FOI_0_tex}
	\centering 
	\begin{tabular}{c|c|c|c|c|c|c|c} 	\hline
		$\omega_r\;(rad/s)$     & $\xi$ & $K_p$   & $K_{r_1}$ & $K_{r_2}$  & $t_s \;(s)$ & $n_s$  & $M_o\;(\%)$  \\ \hline  
		$0.1\,\omega_u = 0.132$ & $0$   & $1.01$  & $0.0699$  & $-0.0174$  & $125.7$     & $2.6$  & $9.9$    \\ \hline 
		$0.7\,\omega_u = 0.924$ & $0$   & $0.524$ & $0.0120$  & $-0.443$   & $58$        & $8.5$ & $23$ \\ \hline 
	\end{tabular} 
\end{table}

\begin{figure}[t!]
	\centering
	\subfigure[$\omega_r = 0.1\,\omega_u$ - The Nyquist diagram of $C_{pr(s)}G_a(s)$ has two turns with radius tending to infinity in the right half of the complex plane.]{\includegraphics[width=0.8\columnwidth]{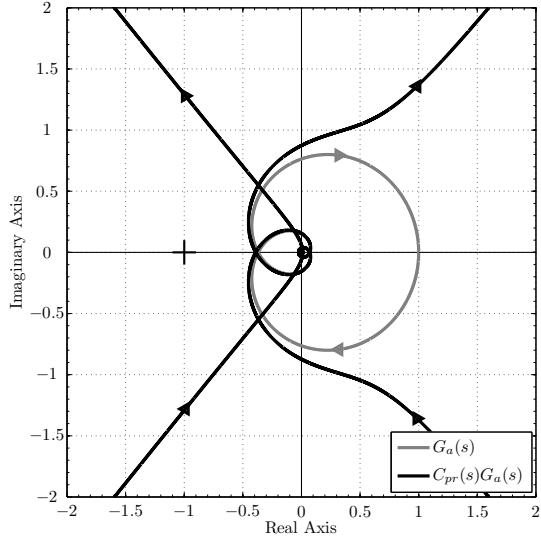}
		\label{fig:nyquist_G_d_0ze_2pe_real_1_wu_0_1_z_cont_res_FOI_0}}
		\vskip-0.18cm
	\subfigure[$\omega_r = 0.7\,\omega_u$ - The Nyquist diagram of $C_{pr(s)}G_a(s)$ has two turns with radius tending to infinity in the right half of the complex plane.]{\includegraphics[width=0.8\columnwidth]{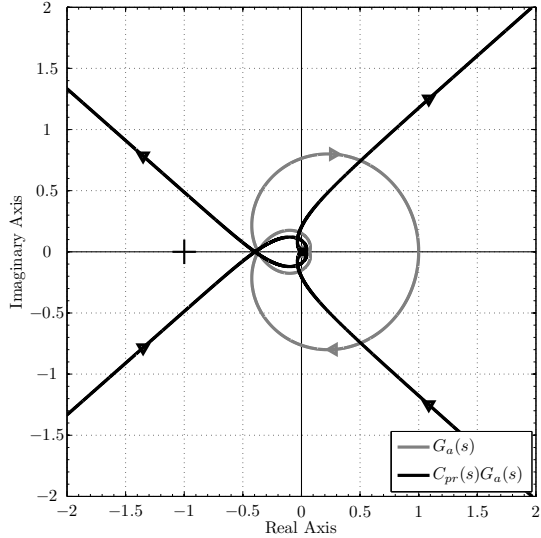}
		\label{fig:nyquist_G_d_0ze_2pe_real_1_wu_0_7_cont_res_FOI_0}}
	\caption{Nyquist diagrams $G_a(s)$ and $C_{pr}(s)G_a(s)$.}
	\label{fig:nyquist_G_d_0ze_2pe_real_1}
\end{figure}

A frequency response analysis is useful to analyze the system stability. Nyquist diagrams of the plant's transfer function $G_a(s)$ and of the loop transfer function $C_{pr(s)}G_a(s)$ 
are presented in \figref{fig:nyquist_G_d_0ze_2pe_real_1_wu_0_1_z_cont_res_FOI_0} for $\omega_{r} = 0.1\,\omega_{u}$ and in \figref{fig:nyquist_G_d_0ze_2pe_real_1_wu_0_7_cont_res_FOI_0} for $\omega_{r} = 0.7\,\omega_{u}$, where it can be seen that for both reference frequencies the Nyquist diagrams of $C_{pr(s)}G_a(s)$ do not encircle the point $-1+j0$. For $\omega_{r} = 0.1\,\omega_{u}$, the frequency response is smooth enough around the negative real axis so that shifting the ultimate point away from $-1+j0$ guarantees good stability margins. Unlike for $\omega_{r} = 0.7\,\omega_{u}$, where proximity of $\omega_r$ and $\omega_u$ implies that the distance between the loop transfer function and $-1+j0$ is decreased. In this case, shifting the plant's ultimate point away from $-1+j0$ does not guarantee good stability margins, since the nearby points are not shifted along because the frequency response is not sufficiently smooth in this range of frequencies. As a consequence, stability margins are much smaller than the case $\omega_{r} = 0.1\,\omega_{u}$ and, therefore, poorer transient response is expected for reference frequencies nearest to the plant's ultimate point. Recall that this is a limitation
imposed by the controller structure and not by the tuning method.

\subsection{Class B plant}\label{sec:plants_b}

The second plant considered has the following transfer function:
\begin{equation}
\label{eq:G_c}
G_b(s) {} = {} \frac{1}{\left(s + 1\right)^2}.
\end{equation}

\begin{figure}[t!]
	\centering
	\includegraphics[width=0.9\columnwidth]{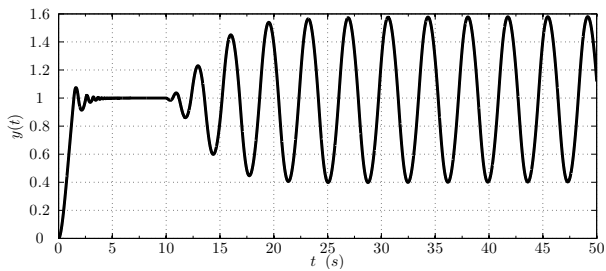}
	\caption{Closed loop response for the RAP experiment applied to the plant $G_b(s)$.}
	\label{fig:osc_G_2_alpha_1}
\end{figure}

The plant described by $G_b(s)$ belongs to the Class B, since it has no ultimate point and its frequency response does cross the $-120\degree$ phase line. An oscillatory condition is obtained for the RAP experiment with $-60\degree$, then it is identified the point of the plant's frequency response whose phase is $\nu = -120\degree$. In \figref{fig:osc_G_2_alpha_1} it is shown the plant's output in this experiment considering as reference input a step with amplitude one, where from $0$ to $10\,s$ it is performed the RAP experiment with $0\degree$, and self-oscillatory behavior is not obtained. Thus, at $10\,s$ the relay's phase is changed to $-60\degree$ and the self-oscillation condition is achieved, as expected. The parameters obtained from this experiment are summarized in \tableref{tab:resultados_G_2_FOI_0_Mu_Tu_tex}.

\begin{table}[t!]
	\centering
	\renewcommand{\tabcolsep}{4pt}
	\caption{Parameters for the RAP experiment and $G_{b}(s)$}
	\label{tab:resultados_G_2_FOI_0_Mu_Tu_tex}
	\centering 
	\begin{tabular}{c|c|c|c|c|c|c|c} 
		\hline
		$d$   & $b$ & $A$     & $\gamma$     & $\nu$         & $|F(j\omega_{120})|$ & $M_{120}$ & $\omega_{120}\;(rad/s)$  \\ 
		\hline 
		$2.4$ & $0$ & $0.589$ & $-60\degree$ & $-120\degree$ & $0.757$              & $0.255$   & $1.69$                 \\ \hline 
	\end{tabular} 
\end{table}

\begin{figure}[t!]
	\centering
	\subfigure[$\omega_r = 0.1\,\omega_u$]{\includegraphics[width=0.85\columnwidth]{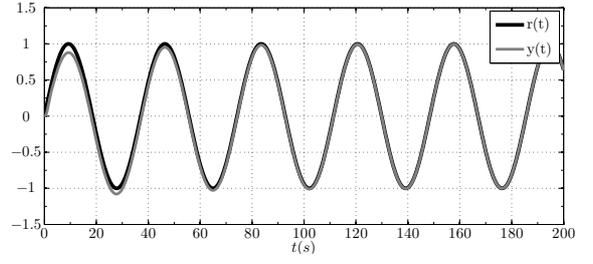}
		\label{fig:out_res_G2_w_0_1}}
	\vskip-0.18cm
	\subfigure[$\omega_r = 0.9\,\omega_u$]{\includegraphics[width=0.85\columnwidth]{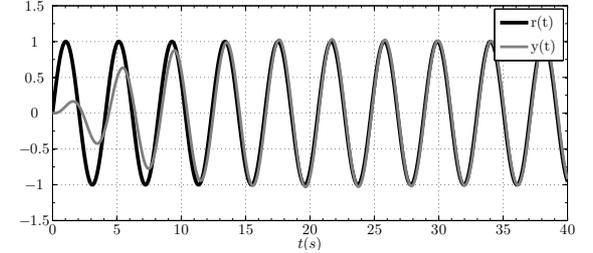}
		\label{fig:out_res_G2_w_0_9}}
	\caption{Closed-loop response of $G_b(s)$ with PR controller.}
	\label{fig:out_res_G2}
\end{figure}

The controller gains are calculated from \eqref{eq:Cpr_class_b_xi} considering $\xi = 0$ and two frequencies: $\omega_r = 0.1\,\omega_{120}$, and  $\omega_r = 0.9\,\omega_{120}$. The sets of controllers' parameters and performance measures are summarized in \tableref{tab:resultados_G_2_FOI_0_tex}. The reference and the output signals for each set of controllers' parameters are shown in \figref{fig:out_res_G2}. 

\begin{table}[t!]
	\centering
	\renewcommand{\tabcolsep}{3pt}
	\caption{Tuning and performance parameters for $G_{b}(s)$}
	\label{tab:resultados_G_2_FOI_0_tex}
	\centering 
	\begin{tabular}{c|c|c|c|c|c|c|c} 	\hline
		$\omega_r\;(rad/s)$         & $\xi$ & $K_p$   & $K_{r_1}$ & $K_{r_2}$ & $t_s \;(s)$ & $n_s$ & $M_o\;(\%)$ \\ \hline 
		$0.1\,\omega_{120} = 0.169$ & $0$   & $3.82$  & $1.14$    & $-0.108$  & $76.9$      & $2.1$ & $7.9$ \\ \hline 
		$0.9\,\omega_{120} = 1.52$  & $0$   & $0.740$ & $0.220$   & $-1.69$   & $26.3$      & $6.4$ & $3.0$ \\ \hline 
	\end{tabular} 
\end{table}

\begin{figure}[t!]
	\centering
	\subfigure[$\omega_r = 0.1\,\omega_{120}$ - The Nyquist diagram of $C_{pr(s)}G_b(s)$ has two turns with radius tending to infinity in the right half of the complex plane.]{\includegraphics[width=0.80\columnwidth]{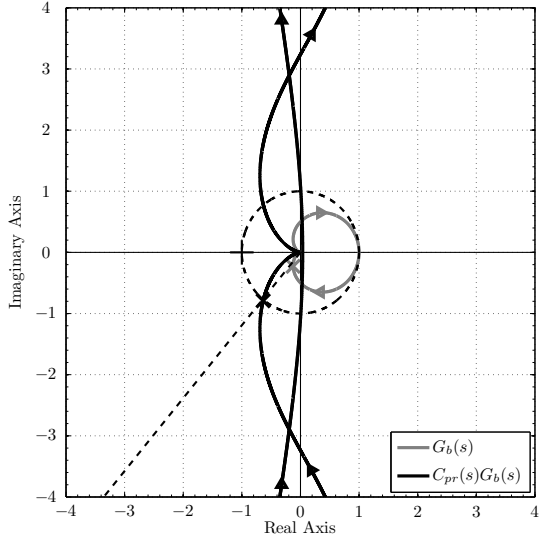}
		\label{fig:nyquist_0ze_2pe_real_a_1_wu_0_1_cont_res_FOI_60}}
	\vskip-0.15cm
	\subfigure[$\omega_r = 0.9\,\omega_{120}$ - The Nyquist diagram of $C_{pr(s)}G_b(s)$ has two turns with radius tending to infinity in the right half of the complex plane.]{\includegraphics[width=0.80\columnwidth]{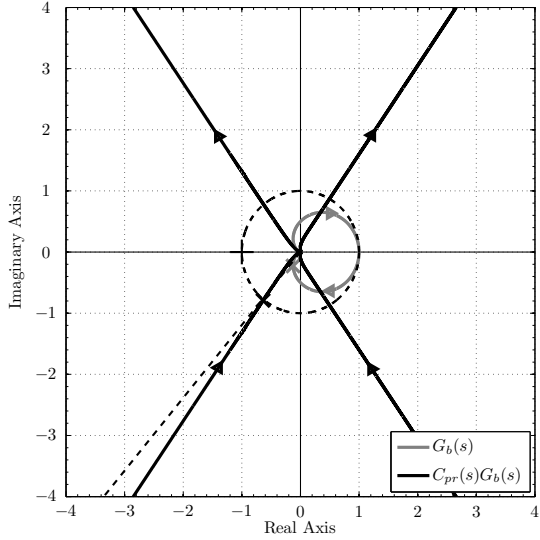}
		\label{fig:nyquist_0ze_2pe_real_a_1_wu_0_9_cont_res_FOI_60}}
	\caption{Nyquist diagrams $G_b(s)$ and $C_{pr}(s)G_b(s)$. Dashed lines are at $-130\degree$ and at unitary magnitude, the identified point of $G_b(s)$ is marked an x.}
	\label{fig:nyquist_0ze_2pe_real_a_1}
\end{figure}

The Nyquist diagram of $G_b(s)$ and of the loop transfer function $C_{pr}(s)G_{b}(s)$ for both $\omega_r$ are shown in \figref{fig:nyquist_0ze_2pe_real_a_1}. For both reference frequencies the Nyquist diagrams of $C_{pr(s)}G_b(s)$ do not encircle the point $-1+j0$ since the frequency response  of the loop transfer function is smooth enough around the negative real axis. As desired, the controller's magnitude and phase at the plant's frequency $\omega_{120}$, respectively, $1/M_{120}$ and $-10\degree$, guarantee phase margin of $50\degree$. 

\subsection{Class C plant}\label{sec:plants_c}

The third plant considered is described by the following transfer function:
\begin{equation} 
\label{eq:G_d}
G_c(s) {} = {} \frac{1}{s + 1}.
\end{equation}

\begin{figure}[t!]
	\centering
	\includegraphics[width=0.9\columnwidth]{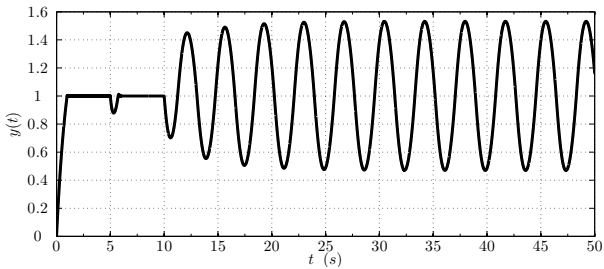}
	\caption{Closed loop response for the RAP experiment applied to the plant $G_c(s)$.}
	\label{fig:osc_G_1_alpha_1}
\end{figure}

The plant $G_c(s)$ has no ultimate point and its frequency response does not cross the $-120\degree$ phase line but it does cross the $-60\degree$ phase line, thus it belongs to the Class C. A self-oscillatory behavior is obtained for the RAP experiment with $-120\degree$, thus the point of the plant's frequency response whose phase is $\nu = -60\degree$ is identified. \figref{fig:osc_G_1_alpha_1} presents the plant's output considering as reference input a step with amplitude one, where from $0$ to $5\,s$ it is implemented the RAP experiment with $0\degree$, at $5\,s$ the relay's phase is changed to $-60\degree$, and self-oscillation condition is not observed. Therefore, at $10\,s$ the RAP experiment with $-120\degree$ is performed and a self-oscillatory behavior is achieved. The parameters obtained from this experiment are summarized in \tableref{tab:resultados_G_2_FOI_0_Mu_Tu_tex}.

\begin{table}[t!]
	\centering
	\renewcommand{\tabcolsep}{4pt}
	\caption{Parameters for the RAP experiment and $G_{c}(s)$}
	\label{tab:resultados_G_1_FOI_0_Mu_Tu_tex}
	\centering 
	\begin{tabular}{c|c|c|c|c|c|c|c} 
		\hline
		$d$   & $b$ & $A$     & $\gamma$      & $\nu$        & $|F(j\omega_{60})|$ & $M_{60}$ & $\omega_{60}\;(rad/s)$  \\ 
		\hline 
		$1.6$ & $0$ & $0.532$ & $-120\degree$ & $-60\degree$ & $0.522$             & $0.500$   & $1.68$  \\ \hline 
	\end{tabular} 
\end{table}

\begin{figure}[t!]
	\centering
	\subfigure[$\omega_r = 0.1\,\omega_u$]{\includegraphics[width=0.85\columnwidth]{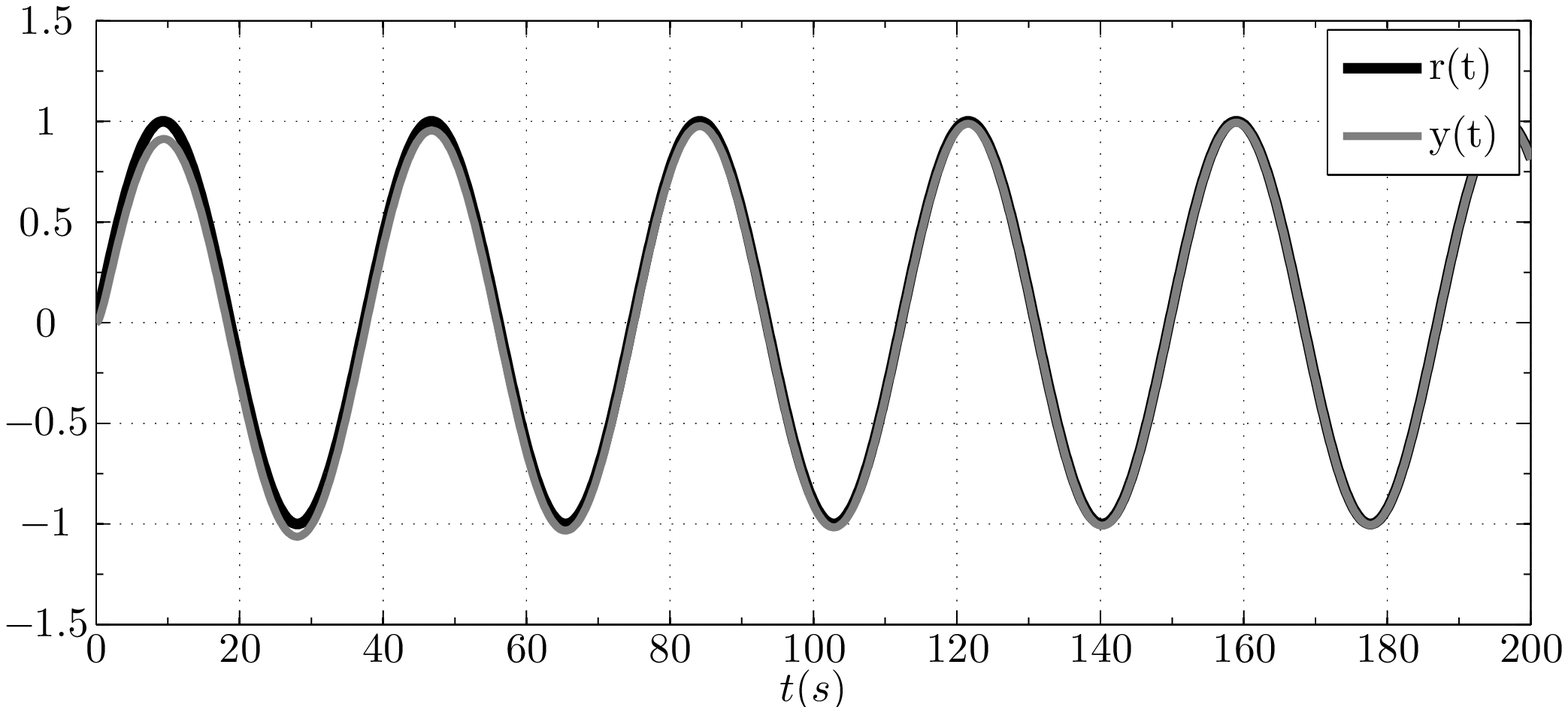}
		\label{fig:out_res_G1_w_0_1}}
	\vskip-0.18cm
	\subfigure[$\omega_r = 0.9\,\omega_u$]{\includegraphics[width=0.85\columnwidth]{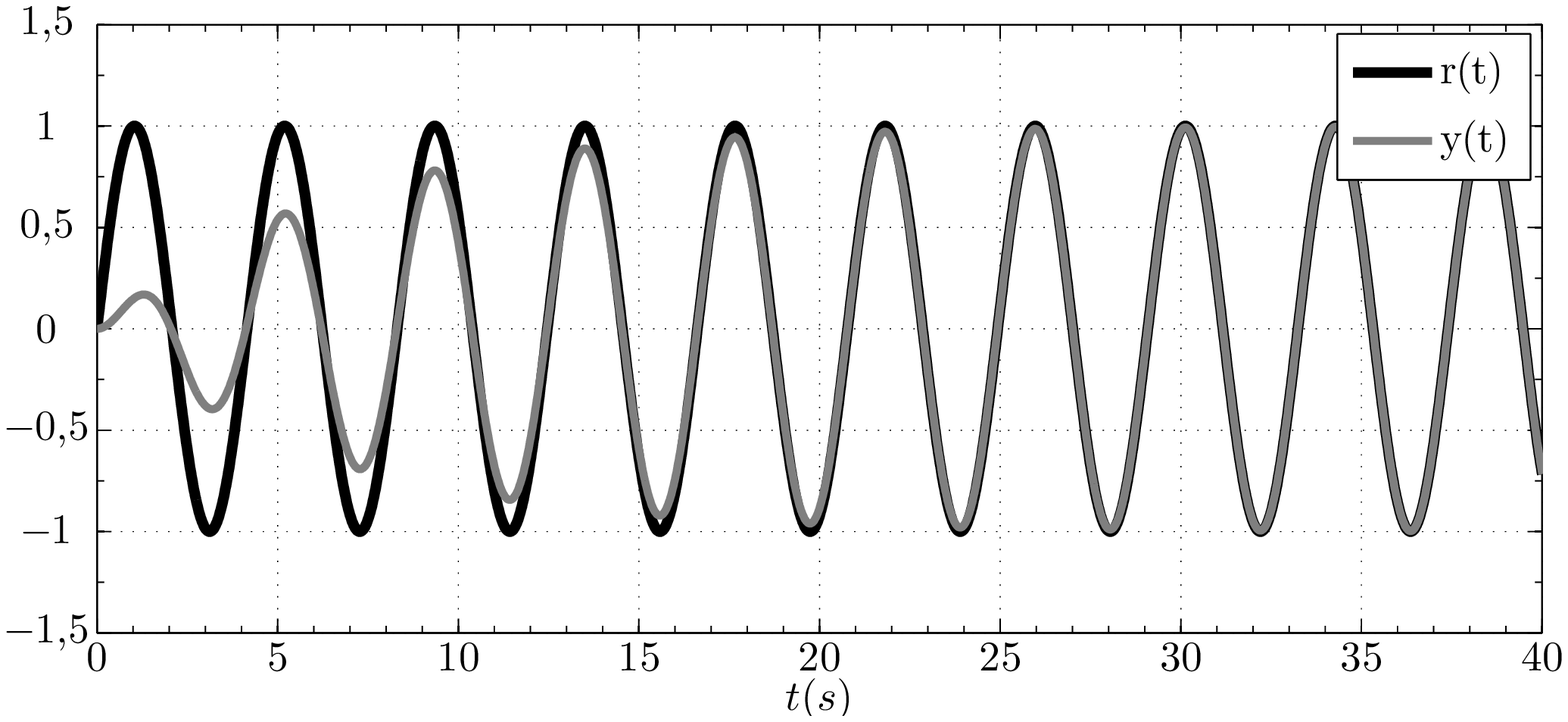}
		\label{fig:out_res_G1_w_0_9}}
	\caption{Closed-loop response of $G_c(s)$ with PR controller.}
	\label{fig:out_res_G1}
\end{figure}

The controller gains are calculated from \eqref{eq:Cpr_class_c_xi_90} using $\xi = 0$ and two frequencies: $\omega_r = 0.1\,\omega_{60}$, and  $\omega_r = 0.9\,\omega_{60}$. The sets of controllers' parameters and performance measures are summarized in \tableref{tab:resultados_G_1_FOI_0_tex}. The reference and the plant's output signals for each set of controllers' parameters are shown in \figref{fig:out_res_G1}.

\begin{table}[t!]
	\centering
	\renewcommand{\tabcolsep}{3pt}
	\caption{Tuning and performance parameters for $G_{c}(s)$}
	\label{tab:resultados_G_1_FOI_0_tex}
	\centering 
	\begin{tabular}{c|c|c|c|c|c|c|c} 	\hline
		$\omega_r\;(rad/s)$        & $\xi$ & $K_p$   & $K_{r_1}$ & $K_{r_2}$ & $t_s \;(s)$ & $n_s$ & $M_o\;(\%)$ \\ \hline 
		$0.1\,\omega_{60} = 0.168$ & $0$   & $1.71$  & $1.66$    & $-0.0479$ & $93.9$      & $2.5$ & $6.3$ \\ \hline 
		$0.9\,\omega_{60} = 1.51$  & $0$   & $0.332$ & $0.319$   & $-0.751$  & $24.1$      & $5.8$ & $0$ \\ \hline 
	\end{tabular} 
\end{table}

\begin{figure}[t!]
	\centering
	\subfigure[$\omega_r = 0.1\,\omega_{60}$ - The Nyquist diagram of $C_{pr(s)}G_c(s)$ has two turns with radius tending to infinity in the right half of the complex plane.]{\includegraphics[width=0.80\columnwidth]{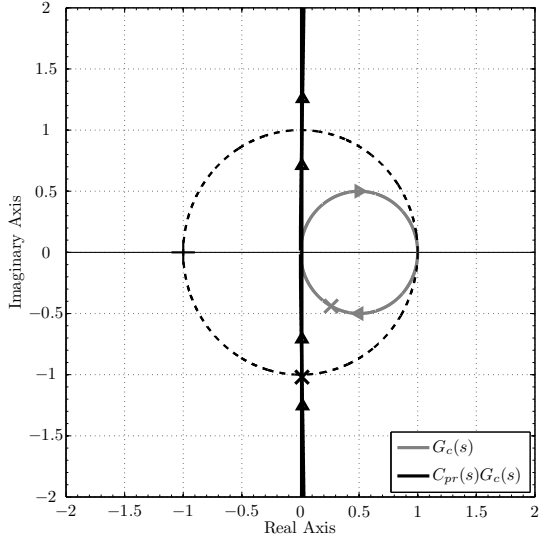}
		\label{fig:nyquist_G_0ze_1pe_a_1_wu_0_1_cont_res_FOI_120}}
	\vskip-0.15cm
	\subfigure[$\omega_r = 0.9\,\omega_{60}$ - The Nyquist diagram of $C_{pr(s)}G_c(s)$ has two turns with radius tending to infinity in the right half of the complex plane.]{\includegraphics[width=0.80\columnwidth]{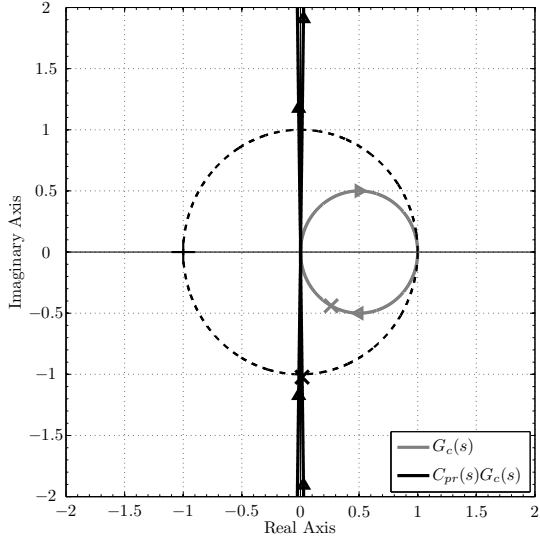}
		\label{fig:nyquist_G_0ze_1pe_a_1_wu_0_9_cont_res_FOI_120}}
	\caption{Nyquist diagrams $G_c(s)$ and $C_{pr}(s)G_c(s)$. Dashed line is at unitary magnitude, the identified point of $G_c(s)$ is marked an x.}
	\label{fig:nyquist_G_0ze_1pe_a_1}
\end{figure}

The corresponding Nyquist diagrams are presented in \figref{fig:nyquist_G_0ze_1pe_a_1}, where it can be seen that the diagram of $C_{pr(s)}G_c(s)$ do not encircle the point $-1+j0$. The controller's magnitude and phase at the plant's frequency $\omega_{60}$, respectively, $1/M_{60}$ and $-30\degree$, guarantee phase margin of $90\degree$. 


\section{Test batch}\label{sec:test_batch}

The proposed methodology has also been applied for tuning of the PR controller in the following classes of plants\footnote{Other fourteen classes of plants with different model parameters were considered in the development of the GFO method, but just few results are presented in this paper for lack of space.} from the test batch proposed in \cite{art:bazanella:2017:PID-rele-foi} and \cite{book:pid:astrom:1995:pid}:
\begin{equation}
\label{eq:G_d1_0ze_2pe_real_dif}
G_{1}(s) {} = {} \frac{e^{-sL}}{\left(Ts + 1\right)\left(T_1s + 1\right)},
\end{equation}
\begin{equation}
\label{eq:G_0ze_3pe_1_2_real_b}
\begin{aligned}
G_{2}(s) {} = {} \frac{1}{\left(s + 1\right)\left(Ts + 1\right)^2},
\end{aligned}
\end{equation}
\begin{equation}
\label{eq:G_0ze_2pe_complex_b}
G_3(s) {} = {} \frac{1}{s + 2 \alpha s + 1},
\end{equation}
\begin{equation}
\label{eq:G_0ze_1pe_b}
G_4(s) {} = {} \frac{\alpha}{s + \alpha}.
\end{equation}

Different values of model parameters ($L$, $T$, $T_1$ and $\alpha$) have been considered for these classes of plants, and tests have been made for different frequencies of the sinusoidal reference $\omega_r$. As the previous examples, initially the RAP experiment has been performed for each of plant, then the closed-loop response to a sinusoidal reference has been evaluated to assess the performance. 

For the plant of Class A represented by the transfer function $G_{1}(s)$ with $T = 1$, $L = T_1 = 0.5$, the ultimate frequency and magnitude are $\omega_u = 2.27$ rad/s and $M_u = 0.282$. The results are shown in \tableref{tab:resultados_G_d1_0ze_2pe_real_dif_T1_0_5_FOI_0_c3_r_0_1_p1_tex}. 
Another set of test was realized for $G_{1}(s)$ considering  $T = 10$, $L = T_1 = 0.5$, which leaded to $\omega_u = 1.75$ rad/s and $M_u = 0.0444$. In this case, the results are summarized in \tableref{tab:resultados_G_d1_0ze_2pe_real_dif_T10_0_5_FOI_0_c3_r_0_1_p1_tex}.

\begin{table}[t!]
	\centering
	\renewcommand{\tabcolsep}{3pt}
	\caption{Tuning and performance for $G_{1}(s)$ with $T = 1$, $L = T_1 = 0.5$}
	\label{tab:resultados_G_d1_0ze_2pe_real_dif_T1_0_5_FOI_0_c3_r_0_1_p1_tex}
	\begin{tabular}{c|c|c|c|c|c|c|c} 
		\hline
		$\omega_r\;(rad/s)$ & $\xi$ &$K_p$ & $K_{r_1}$ & $K_{r_2}$ & $t_s\;(s)$ & $n_s$ & $M_o\;(\%)$ \\ 
		\hline 
		$0.1 \,\omega_u = 0.227 $ & $0$ & $1.4$ & $0.166$ & $-0.0714$ & $60$ & $2.2$ & $11$ \\ \hline 
		$0.3 \,\omega_u = 0.681 $ & $0$ & $1.29$ & $0.153$ & $-0.591$ & $18$ & $2$ & $5.6$ \\ \hline 
		$0.5 \,\omega_u = 1.13 $ & $0$ & $1.07$ & $0.0421$ & $-1.36$ & $9.9$ & $1.8$ & $6.8$ \\ \hline 
		$0.7 \,\omega_u = 1.59 $ & $0$ & $0.726$ & $0.0286$ & $-1.81$ & $37$ & $9.3$ & $27$ \\ \hline 
		$0.9 \,\omega_u = 2.04 $ & $0$ & $0.272$ & $0.0107$ & $-1.12$ & $405$ & $132$ & $70$ \\ \hline 
	\end{tabular} 
\end{table}

\begin{table}[t!]
	\centering
	\renewcommand{\tabcolsep}{3pt}
	\caption{Tuning and performance for $G_{1}(s)$ with $T = 10$, $L = T_1 = 0.5$}
	\label{tab:resultados_G_d1_0ze_2pe_real_dif_T10_0_5_FOI_0_c3_r_0_1_p1_tex}
	\begin{tabular}{c|c|c|c|c|c|c|c} 
		\hline
		$\omega_r\;(rad/s)$ & $\xi$ &$K_p$ & $K_{r_1}$ & $K_{r_2}$ & $t_s\;(s)$ & $n_s$ & $M_o\;(\%)$ \\ 
		\hline 
		$0.1 \,\omega_u = 0.175 $ & $0$ & $8.89$ & $0.816$ & $-0.271$ & $57$ & $1.6$ & $7.7$ \\ \hline 
		$0.3 \,\omega_u = 0.526 $ & $0$ & $8.18$ & $0.75$ & $-2.24$ & $9.3$ & $0.77$ & $3.9$ \\ \hline 
		$0.5 \,\omega_u = 0.877 $ & $0$ & $6.77$ & $0.207$ & $-5.15$ & $20$ & $2.8$ & $18$ \\ \hline 
		$0.7 \,\omega_u = 1.23 $ & $0$ & $4.61$ & $0.14$ & $-6.88$ & $63$ & $12$ & $38$ \\ \hline 
		$0.9 \,\omega_u = 1.58 $ & $0$ & $1.72$ & $0.0523$ & $-4.25$ & $564$ & $142$ & $72$ \\ \hline 
	\end{tabular} 
\end{table}

For the plant of Class A represented by the transfer function $G_{2}(s)$ two sets of tests were performed with $T = 0.05$ and $T = 5$. For $T = 0.05$ were identified $\omega_u = 20.5$ rad/s and $M_u = 0.0234$, and $\omega_u = 0.650$ rad/s and $M_u = 0.0718$ were obtained for $T = 5$. The results are presented in \tableref{tab:resultados_G_0ze_3pe_1_2_real_0_05_FOI_0_c3_r_0_1_p1_tex} and \ref{tab:resultados_G_0ze_3pe_1_2_real_5_FOI_0_c3_r_0_1_p1_tex}.

\begin{table}[t!]
	\centering
	\renewcommand{\tabcolsep}{3pt}
	\caption{Tuning and performance for $G_{2}(s)$ with $T = 0.05$}
	\label{tab:resultados_G_0ze_3pe_1_2_real_0_05_FOI_0_c3_r_0_1_p1_tex}
	\begin{tabular}{c|c|c|c|c|c|c|c} \hline
		$\omega_r\;(rad/s)$ & $\xi$ &$K_p$ & $K_{r_1}$ & $K_{r_2}$ & $t_s\;(s)$ & $n_s$ & $M_o\;(\%)$ \\ \hline 
		$0.1 \,\omega_u = 2.05 $ & $0$ & $16.9$ & $18.2$ & $-70.6$ & $6.2$ & $2$ & $6$ \\ \hline 
		$0.3 \,\omega_u = 6.16 $ & $0$ & $15.5$ & $16.7$ & $-584$ & $1.5$ & $1.5$ & $1.9$ \\ \hline 
		$0.5 \,\omega_u = 10.3 $ & $0$ & $12.8$ & $4.6$ & $-1343$ & $2.4$ & $4$ & $25$ \\ \hline 
		$0.7 \,\omega_u = 14.4 $ & $0$ & $8.76$ & $3.13$ & $-1794$ & $7.4$ & $17$ & $47$ \\ \hline 
		$0.9 \,\omega_u = 18.5 $ & $0$ & $3.27$ & $1.16$ & $-1108$ & $70$ & $205$ & $78$ \\ \hline 
	\end{tabular} 
\end{table}

\begin{table}[t!]
	\centering
	\renewcommand{\tabcolsep}{3pt}
	\caption{Tuning and performance for $G_{2}(s)$ with $T = 5$}
	\label{tab:resultados_G_0ze_3pe_1_2_real_5_FOI_0_c3_r_0_1_p1_tex}
	\begin{tabular}{c|c|c|c|c|c|c|c} \hline
		$\omega_r\;(rad/s)$ & $\xi$ &$K_p$ & $K_{r_1}$ & $K_{r_2}$ & $t_s\;(s)$ & $n_s$ & $M_o\;(\%)$ \\ \hline 
		$0.1 \,\omega_u = 0.065 $ & $0$ & $5.5$ & $0.187$ & $-0.023$ & $126$ & $1.3$ & $12$ \\ \hline 
		$0.3 \,\omega_u = 0.195 $ & $0$ & $5.06$ & $0.172$ & $-0.19$ & $68$ & $2.1$ & $5$ \\ \hline 
		$0.5 \,\omega_u = 0.325 $ & $0$ & $4.19$ & $0.0474$ & $-0.437$ & $82$ & $4.2$ & $26$ \\ \hline 
		$0.7 \,\omega_u = 0.455 $ & $0$ & $2.85$ & $0.0322$ & $-0.584$ & $241$ & $17$ & $48$ \\ \hline 
		$0.9 \,\omega_u = 0.585 $ & $0$ & $1.07$ & $0.012$ & $-0.361$ & $2262$ & $210$ & $78$ \\ \hline 
	\end{tabular} 
\end{table}

For the plant of Class B represented by the transfer function $G_{3}(s)$ also two sets of tests were performed with $\alpha = 0.1$ and $\alpha = 0.7$. For $\alpha = 0.1$, the plant's magnitude and frequency at the point of $\nu = -120\degree$ are, respectively, $\omega_{120} = 1.06$ rad/s and $M_{120} = 4.08$, whereas for $\alpha = 0.7$ the identified quantities are $\omega_{120} = 1.46$ rad/s and $M_{120} = 0.422$. In these cases, the results are presented in \tableref{tab:resultados_G_0ze_2pe_complex_0_1_FOI_60_c3_r_0_1_tex_tex} and \ref{tab:resultados_G_0ze_2pe_complex_0_7_FOI_60_c3_r_0_1_tex}.

\begin{table}[t!]
	\centering
	\renewcommand{\tabcolsep}{2.5pt}
	\caption{Tuning and performance for $G_{3}(s)$ with $\alpha = 0.1$}
	\label{tab:resultados_G_0ze_2pe_complex_0_1_FOI_60_c3_r_0_1_tex_tex}
	\begin{tabular}{c|c|c|c|c|c|c|c} \hline
		$\omega_r\;(rad/s)$ & $\xi$ &$K_p$ & $K_{r_1}$ & $K_{r_2}$ & $t_s\;(s)$ & $n_s$ & $M_o\;(\%)$ \\ \hline 
		$0.1 \,\omega_{120} = 0.106 $ & $0$ & $0.239$ & $0.0447$ & $-0.00265$ & $192$ & $3.2$ & $2$ \\ \hline 
		$0.3 \,\omega_{120} = 0.318 $ & $0$ & $0.22$ & $0.0411$ & $-0.022$ & $193$ & $9.8$ & $16$ \\ \hline 
		$0.5 \,\omega_{120} = 0.53 $ & $0$ & $0.181$ & $0.0339$ & $-0.0504$ & $168$ & $14$ & $24$ \\ \hline 
		$0.7 \,\omega_{120} = 0.741 $ & $0$ & $0.124$ & $0.023$ & $-0.0673$ & $133$ & $16$ & $24$ \\ \hline 
		$0.9 \,\omega_{120} = 0.953 $ & $0$ & $0.0462$ & $0.00858$ & $-0.0416$ & $76$ & $12$ & $13$ \\ \hline 
	\end{tabular} 
\end{table}

\begin{table}[t!]
	\centering
	\renewcommand{\tabcolsep}{3pt}
	\caption{Tuning and performance for $G_{3}(s)$ with $\alpha= 0.7$}
	\label{tab:resultados_G_0ze_2pe_complex_0_7_FOI_60_c3_r_0_1_tex}
	\begin{tabular}{c|c|c|c|c|c|c|c} \hline
		$\omega_r\;(rad/s)$ & $\xi$ &$K_p$ & $K_{r_1}$ & $K_{r_2}$ & $t_s\;(s)$ & $n_s$ & $M_o\;(\%)$ \\ \hline 
		$0.1 \,\omega_{120} = 0.146 $ & $0$ & $2.31$ & $0.597$ & $-0.0489$ & $66$ & $1.5$ & $9.3$ \\ \hline 
		$0.3 \,\omega_{120} = 0.439 $ & $0$ & $2.13$ & $0.549$ & $-0.405$ & $24$ & $1.6$ & $7.4$ \\ \hline 
		$0.5 \,\omega_{120} = 0.731 $ & $0$ & $1.76$ & $0.452$ & $-0.928$ & $18$ & $2.1$ & $7.7$ \\ \hline 
		$0.7 \,\omega_{120} = 1.02 $ & $0$ & $1.2$ & $0.307$ & $-1.24$ & $11$ & $1.8$ & $6.3$ \\ \hline 
		$0.9 \,\omega_{120} = 1.32 $ & $0$ & $0.447$ & $0.115$ & $-0.766$ & $30$ & $6.3$ & $3.2$ \\ \hline 
	\end{tabular} 
\end{table}

Finally, for the plant of Class C described by the transfer function $G_{4}(s)$ also two sets of tests were performed with $\alpha = 0.1$ and $\alpha = 100$. For $\alpha = 0.1$, the plant's magnitude and frequency at the point of $\nu = -60\degree$ are, respectively, $\omega_{60} = 0.169$ rad/s and $M_{60} = 0.498$, whereas for $\alpha = 100$, $\omega_{60} = 168$ rad/s and $M_{60} = 0.500$ are obtained. In these cases, the results are summarized in Tables \ref{tab:resultados_G_0ze_1pe_0_1_FOI_120_c3_r_0_1_tex} and \ref{tab:resultados_G_0ze_1pe_100_FOI_120_c3_r_0_1_tex}.
	
\begin{table}[t!]
	\centering
	\renewcommand{\tabcolsep}{2.5pt}
	\caption{Tuning and performance for $G_{4}(s)$ with $\alpha = 0.1$}
	\label{tab:resultados_G_0ze_1pe_0_1_FOI_120_c3_r_0_1_tex}
	\begin{tabular}{c|c|c|c|c|c|c|c} \hline
		$\omega_r\;(rad/s)$ & $\xi$ &$K_p$ & $K_{r_1}$ & $K_{r_2}$ & $t_s\;(s)$ & $n_s$ & $M_o\;(\%)$ \\ \hline 
		$0.1 \,\omega_{60} = 0.0169$ & 0 & $1.72$ & $0.167$ & $-0.000484$ & 937& $2.5$ & $6.3$ \\ \hline 
		$0.3 \,\omega_{60} = 0.0506$ & 0 & $1.58$ & $0.154$ & $-0.00401$ &159& $1.3$ & $7.3$ \\ \hline 
		$0.5 \,\omega_{60} = 0.0843$ & 0 & $1.31$ & $0.127$ & $-0.00919$ &50.3& $0.68$ & $0.48$ \\ \hline 
		$0.7 \,\omega_{60} = 0.118$ & 0 & $0.891$ & $0.0863$ & $-0.0123$ &80.4& $1.5$ & $0$ \\ \hline 
		$0.9 \,\omega_{60} = 0.152$ & 0 & $0.333$ & $0.0321$ & $-0.00758$ &240& $5.8$ & $0$ \\ \hline 
	\end{tabular} 
\end{table}

	\begin{table}[t!]
		\centering
		\renewcommand{\tabcolsep}{3pt}
		\caption{Tuning and performance for $G_{4}(s)$ with $\alpha= 100$}
		\label{tab:resultados_G_0ze_1pe_100_FOI_120_c3_r_0_1_tex}
		\begin{tabular}{c|c|c|c|c|c|c|c} 
			\hline
			$\omega_r\;(rad/s)$ & $\xi$ &$K_p$ & $K_{r_1}$ & $K_{r_2}$ & $t_s\;(s)$ & $n_s$ & $M_o\;(\%)$ \\ \hline 
			$0.1 \,\omega_{60} = 16.8 $ & $0$ & $1.71$ & $166$ & $-476$ & $0.94$ & $2.5$ & $6.3$ \\ \hline 
			$0.3 \,\omega_{60} = 50.3 $ & $0$ & $1.58$ & $152$ & $-3942$ & $0.16$ & $1.3$ & $7.3$ \\ \hline 
			$0.5 \,\omega_{60} = 83.8 $ & $0$ & $1.3$ & $126$ & $-9040$ & $0.051$ & $0.67$ & $0.5$ \\ \hline 
			$0.7 \,\omega_{60} = 117 $ & $0$ & $0.887$ & $85.4$ & $-12078$ & $0.081$ & $1.5$ & $0$ \\ \hline 
			$0.9 \,\omega_{60} = 151 $ & $0$ & $0.331$ & $31.8$ & $-7462$ & $0.24$ & $5.8$ & $0$ \\ \hline 
		\end{tabular} 
	\end{table}

It is observed in the results presented in these last two sections that:
\begin{enumerate}
	\item For most cases good performance and robustness was obtained.
	The resulting transient performance is similar to what is typically achieved with the celebrated
	CFO method for PI tuning. 
	\item For the Class A, performance deteriorates as the reference frequency approaches the plant's ultimate frequency, as expected from the analysis, since the phase margin is reduced in this situation, but appropriate performance is obtained provided that 
	$\omega_r < 0.9\;\omega_u$. 
	\item For the Classes B and C, performance does not deteriorate so significantly as the reference frequency approaches the identified frequency since the chosen phase margin is guaranteed for both classes of plants, and appropriate performance is obtained 
	for all $\omega_r <0.9 \;\omega_{120}$ (for Class B) or $\omega_r < 0.9\;\omega_{60}$ (for Class C).
\end{enumerate}

\section{Conclusion}\label{sec:conclusion}

This paper has proposed a tuning method for PR controllers including both plants that have an ultimate point and plants that do not.
The methodology is based on the identification of a particularly relevant point of the plant's frequency response through the RAP experiment. Four sets of tuning formulas have been developed to obtain appropriate stability margins for each of the considered classes of plants. The proposed methodology, which includes the identification experiment and the developed tuning formulas, was validated considering a wide variety of plants and also different reference frequencies below the plant's identified frequency. 
For all such cases good performance and robustness have been obtained. Thus, the proposed methodology is an alternative to experimentally tune a PR controller without the plant model and also without the use of advanced control system techniques, that should contribute to the applicability and enlarge the dissemination of resonant controllers. Possible extension of this work is development of the tuning method for proportional multi-resonant controllers. 


\bibliographystyle{IEEETran}   
\bibliography{bibliography}    

\begin{thebibliography}{10}
\providecommand{\url}[1]{#1}
\csname url@samestyle\endcsname
\providecommand{\newblock}{\relax}
\providecommand{\bibinfo}[2]{#2}
\providecommand{\BIBentrySTDinterwordspacing}{\spaceskip=0pt\relax}
\providecommand{\BIBentryALTinterwordstretchfactor}{4}
\providecommand{\BIBentryALTinterwordspacing}{\spaceskip=\fontdimen2\font plus
\BIBentryALTinterwordstretchfactor\fontdimen3\font minus
  \fontdimen4\font\relax}
\providecommand{\BIBforeignlanguage}[2]{{%
\expandafter\ifx\csname l@#1\endcsname\relax
\typeout{** WARNING: IEEEtran.bst: No hyphenation pattern has been}%
\typeout{** loaded for the language `#1'. Using the pattern for}%
\typeout{** the default language instead.}%
\else
\language=\csname l@#1\endcsname
\fi
#2}}
\providecommand{\BIBdecl}{\relax}
\BIBdecl

\bibitem{art:contr:francis:1975}
B.~A. Francis and W.~M. Wonham, ``The internal model principle for linear
  multivariable regulators,'' \emph{Applied Mathematics and Optimization},
  vol.~2, no.~2, pp. 170--194, June 1975.

\bibitem{art:res:per:2014:multiplos_ressonantes}
L.~F.~A. Pereira, J.~V. Flores, G.~Bonan, D.~F. Coutinho, , and J.~M. {Gomes da
  Silva Jr.}, ``Multiple resonant controllers for uninterruptible power
  supplies -- {A} systematic robust control design approach,'' \emph{{IEEE}
  Transactions on Industrial Electronics}, vol.~61, no.~3, pp. 1528--1538, Mar
  2014.

\bibitem{art:res:Fukuda:2001:filtro_ativo}
S.~Fukuda and T.~Yoda, ``A novel current-tracking method for active filters
  based on a sinusoidal internal model [for pwm invertors],'' \emph{IEEE
  Transactions on Industry Applications}, vol.~37, no.~3, pp. 888--895, May
  2001.

\bibitem{art:res:Gonzatti:2016:active_filter}
R.~B. Gonzatti, S.~C. Ferreira, C.~H. da~Silva, R.~R. Pereira, L.~E.~B.
  da~Silva, and G.~Lambert-Torres, ``Smart impedance: A new way to look at
  hybrid filters,'' \emph{IEEE Transactions on Smart Grid}, vol.~7, no.~2, pp.
  837--846, 2016.

\bibitem{art:res:Teodorescu:2006:ups}
R.~Teodorescu, F.~Blaabjerg, M.~Liserre, and P.~C. Loh, ``Proportional-resonant
  controllers and filters for grid-connected voltage-source converters,''
  \emph{IEE Proceedings Electric Power Applications}, vol. 153, no.~5, pp.
  750--762, Sept. 2006.

\bibitem{art:res:Habibullah:2017:vibracao}
H.~Habibullah, H.~R. Pota, and I.~R. Petersen, ``A novel control approach for
  high precision positioning of a piezoelectric tube scanner,'' \emph{IEEE
  Transactions on Automation Science and Engineering}, vol.~14, no.~1, pp.
  325--336, Jan 2017.

\bibitem{art:res:Moheimani:2005:ress_struc}
S.~O.~R. Moheimani and B.~J.~G. Vautier, ``Resonant control of structural
  vibration using charge-driven piezoelectric actuators,'' \emph{IEEE
  Transactions on Control Systems Technology}, vol.~13, no.~6, pp. 1021--1035,
  Nov 2005.

\bibitem{art:res:Abosh:2017:torq}
A.~H. Abosh, Z.~Q. Zhu, and Y.~Ren, ``Reduction of torque and flux ripples in
  space vector modulation-based direct torque control of asymmetric permanent
  magnet synchronous machine,'' \emph{IEEE Transactions on Power Electronics},
  vol.~32, no.~4, pp. 2976--2986, April 2017.

\bibitem{art:pereira:2015:PR-ZN}
L.~F.~A. Pereira and A.~S. Bazanella, ``Tuning rules for proportional resonant
  controllers,'' \emph{IEEE Transactions on Control Systems Technology},
  vol.~23, no.~5, pp. 2010--2017, Sept 2015.

\bibitem{art:astrom:1984:rele}
K.~J. {\AA}str{\"o}m and T.~H{\"a}gglund, ``Automatic tuning of simple
  regulators with specifications on phase and amplitude margins,''
  \emph{Automatica}, vol.~20, no.~5, pp. 645--651, Sept. 1984.

\bibitem{art:bazanella:2017:PID-rele-foi}
A.~S. Bazanella, L.~F.~A. Pereira, and A.~Parraga, ``A new method for pid
  tuning including plants without ultimate frequency,'' \emph{{IEEE}
  Transactions on Control Systems Technology}, vol.~25, no.~2, pp. 637--644,
  March 2017.

\bibitem{art:pid:ZN:1942}
J.~G. Ziegler, N.~B. Nichols, and N.~Y. Rochester, ``Optimum settings for
  automatic controllers,'' \emph{Transactions of the ASME}, vol.~64, no.~11,
  pp. 759--768, Nov. 1942.

\bibitem{book:pid:astrom:1995:pid}
K.~J. {\AA}str{\"o}m and T.~H{\"a}gglund, \emph{PID controllers: theory,
  design, and tuning}.\hskip 1em plus 0.5em minus 0.4em\relax Research Triangle
  Park, NC, USA: ISA, 1995.

\bibitem{book:wolovich:1993:automatic}
W.~A. Wolovich, \emph{Automatic control systems: basic analysis and
  design}.\hskip 1em plus 0.5em minus 0.4em\relax New York, NY, USA: Oxford
  University Press, 1993.

\bibitem{art:foi:tepljakov:2011:fomcon}
A.~Tepljakov, E.~Petlenkov, and J.~Belikov, ``{FOMCON}: a {MATLAB} toolbox for
  fractional-order system identification and control,'' \emph{International
  Journal of Microelectronics and Computer Science}, vol.~2, no.~2, pp. 51--62,
  2011.

\bibitem{online:foi:Tepljakov:2016:FOI_site}
\BIBentryALTinterwordspacing
A.~Tepljakov. {FOMCON}: {Fractional-Order Modeling and Control}. Accessed on
  Oct. 2017. [Online]. Available: \url{http://fomcon.net/}
\BIBentrySTDinterwordspacing

\end{thebibliography}
	
\end{document}